\documentclass[twocolumn, 10pt, aps, superscriptaddress, floatfix, showpacs, prb, citeautoscript]{revtex4-1}

\usepackage[utf8]{inputenc}
\usepackage[T1]{fontenc}
\usepackage{graphicx}
\usepackage{amsmath}
\usepackage{amssymb}
\usepackage{wasysym}
\usepackage{bm}
\usepackage{xcolor}
\usepackage[colorlinks, citecolor={blue!50!black}, urlcolor={blue!50!black}, linkcolor={red!50!black}]{hyperref}
\usepackage{bookmark}
\usepackage{tabularx}
\usepackage{mathtools}
\usepackage{microtype}
\usepackage[load=physical,load=abbr]{siunitx}

\newcommand{\co}[2]{#2}

\DeclarePairedDelimiter\abs{\lvert}{\rvert}%
\DeclarePairedDelimiter\norm{\lVert}{\rVert}%
\makeatletter
\let\oldabs\abs
\def\abs{\@ifstar{\oldabs}{\oldabs*}}
\let\oldnorm\norm
\def\norm{\@ifstar{\oldnorm}{\oldnorm*}}
\makeatother

\newcolumntype{L}[1]{>{\raggedright\arraybackslash}p{#1}}
\newcolumntype{C}[1]{>{\centering\arraybackslash}p{#1}}
\newcolumntype{R}[1]{>{\raggedleft\arraybackslash}p{#1}}

\graphicspath{{figures/}}

\begin{document}
\title{The Andreev rectifier: a nonlocal conductance signature of topological phase transitions}

\author{T. \"O. Rosdahl}
\email[Electronic address: ]{torosdahl@gmail.com}
\affiliation{Kavli Institute of Nanoscience, Delft University of Technology,
  P.O. Box 4056, 2600 GA Delft, The Netherlands}
\author{A. Vuik}
\email[Electronic address: ]{adriaanvuik@gmail.com}
\affiliation{Kavli Institute of Nanoscience, Delft University of Technology,
  P.O. Box 4056, 2600 GA Delft, The Netherlands}
 \author{M. Kjaergaard}
 \affiliation{Center for Quantum Devices and Station Q Copenhagen, Niels Bohr Institute, University of Copenhagen, Universitetsparken 5, 2100 Copenhagen, Denmark}
 \affiliation{Research Laboratory of Electronics, Massachusetts Institute of Technology, Cambridge, MA 02139, USA}
\author{A. R. Akhmerov}
\affiliation{Kavli Institute of Nanoscience, Delft University of Technology,
  P.O. Box 4056, 2600 GA Delft, The Netherlands}

\date{October 24, 2017}

\begin{abstract}
The proximity effect in hybrid superconductor-semiconductor structures, crucial for realizing Majorana edge modes, is complicated to control due to its dependence on many unknown microscopic parameters.
In addition, defects can spoil the induced superconductivity locally in the proximitised system which complicates measuring global properties with a local probe.
We show how to use the nonlocal conductance between two spatially separated leads to probe three global properties of a proximitised system: the bulk superconducting gap, the induced gap, and the induced coherence length.
Unlike local conductance spectroscopy, nonlocal conductance measurements distinguish between non-topological zero-energy modes localized around potential inhomogeneities, and true Majorana edge modes that emerge in the topological phase.
In addition, we find that the nonlocal conductance is an odd function of bias at the topological phase transition, acting as a current rectifier in the low-bias limit.
More generally, we identify conditions for crossed Andreev reflection to dominate the nonlocal conductance and show how to design a Cooper pair splitter in the open regime.
\end{abstract}

\maketitle
\section{Introduction}
\label{sec:intro}
\co{The proximity effect induces superconductivity in non-superconducting materials, which among other things gives Majoranas.}
The superconducting proximity effect occurs when a normal material (metal) is placed in contact with a superconductor. The resulting transfer of superconducting properties to the normal material \cite{degennes1964, blonder1982} makes it possible to explore induced superconductivity in a range of materials that are not intrinsically superconducting, for example in ferromagnetic metals \cite{tokuyasu1988, demler1997, gingrich2016} and in graphene \cite{titov2006, heersche2007, calado2015}. Another recent application of the proximity effect is the creation of the Majorana quasiparticle \cite{alicea2012, leijnse2012, beenakker2013}, which is a candidate for the realization of topological quantum computation \cite{bravyi2002}, and a focus of research efforts in recent years \cite{mourik2012, das2012, deng2012}.

\begin{figure}[!tbh]
\includegraphics[width=0.97\columnwidth]{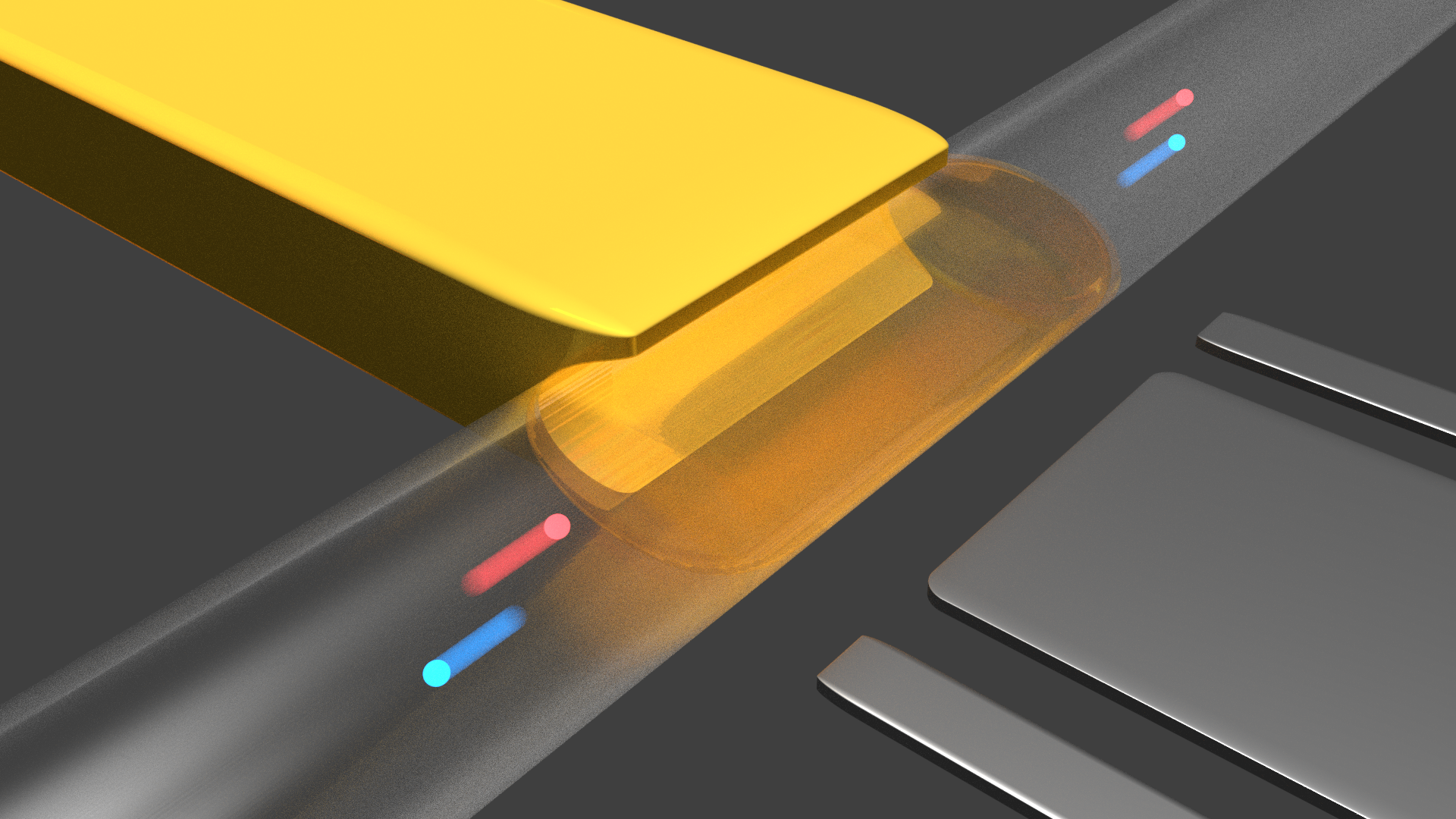}
\caption{A superconductor (yellow) proximitises a semiconducting region (transparent) from the side. Narrow gates control the coupling of the proximitised scattering region with the leads, a wider gate controls the chemical potential. An incoming electron from the left (red dot) either undergoes a local process, i.e.\ Andreev reflection into a hole (blue outgoing dot to the left) or normal reflection (not shown), or a nonlocal process (outgoing electron or hole to the right).}
\label{fig:sketch}
\end{figure}

\co{The quality of induced superconductivity depends on uncontrollable and unknown factors, and at the same time it is hard to probe.}
The proximity effect is due to the Andreev reflection of quasiparticles at the interface with the superconductor \cite{blonder1982}, which forms correlated electron-hole pairs that induce superconductivity in the normal material.
This makes the proximity effect in real systems sensitive to microscopic interface properties, such as coupling strength, charge accumulation and lattice mismatch \cite{klapwijk2004, gueron1996}.
Spatial inhomogeneities in the proximitised system, such as charge defects, may furthermore spoil the induced correlations locally.
In a typical proximity setup, the superconductor proximitises an extended region of a normal material, as shown in Fig. \ref{fig:sketch}.
A normal lead attached  to one of the ends of the proximitised region probes the response to an applied voltage.
When the coupling between the lead and the proximitised region is weak, the lead functions as a tunnel probe of the density of states in the latter.
Since induced superconductivity may be inhomogeneous, and Andreev reflection happens locally, such an experiment only probes the region in direct vicinity of the normal lead, and not the overall properties of the proximitised region.
For example, if the electrostatic potential is inhomogeneous, it may create accidental low-energy modes that are nearly indistinguishable from Majoranas \cite{pikulin2012, Kells2012, mi2014, Prada2012, Moore2016, liu2017}.

\co{The nonlocal response reveals properties of induced superconductivity.}
We show how the \emph{nonlocal} response between two spatially separated normal leads (see Fig.~\ref{fig:sketch}) may be used to probe both the bulk superconducting gap $\Delta$ and the induced gap $\Delta_{\mathrm{ind}}$, as well as the induced coherence length $\xi$.
At subgap energies, quasiparticles propagate as evanescent waves with the decay length $\xi$ in the proximitised system. This suppresses the nonlocal response with increasing separation $L$ between the two normal leads \cite{byers1995, russo2005, reinthaler2013}.
Therefore, the length dependence of the nonlocal conductance measures when two ends of a proximitised system are effectively decoupled.
When $L/\xi \gtrsim 1$,  the nonlocal conductance is only possible in the energy window between the bulk superconducting gap $\Delta$ and the induced gap $\Delta_\text{ind}$.
The sensitivity to an induced gap allows one to use nonlocal conductance to distinguish between a induced gap closing and an Andreev level crossing at zero energy.
In contrast, a local measurement may produce a similar result in both cases.

\co{With Zeeman field, we create a Cooper pair splitter and an Andreev rectifier}
Two processes constitute the nonlocal response: direct electron transfer between the normal leads, and the crossed Andreev reflection (CAR) of an electron from one lead into a hole in the second lead \cite{falci2001, melin2002}.
Experimental \cite{beckmann2004, brauer2010, schindele2014} and theoretical \cite{recher2001, morten2006, haugen2010, crepin2015, zhang2017-2, liu2017-2} studies of CAR-dominated signals aim at producing a Cooper pair splitter \cite{deutscher2000, hofstetter2009, herrmann2010}, which has potential applications in quantum information processing.
We show that applying a Zeeman field in the proximitised system creates wide regions in parameter space where CAR dominates the nonlocal response.
Furthermore, we demonstrate how to obtain a CAR dominated signal in the absence of a Zeeman field in the low-doping regime.
Finally, we prove that at the topological phase transition and with $L/\xi \gtrsim 1$, the nonlocal conductance is an approximately odd function of bias.
This phenomenon only relies on particle-hole symmetry, and hence manifests both in clean and disordered junctions.
Therefore, a proximitised system coupled to normal leads acts as a rectifier of the applied voltage bias universally at the topological phase transition.

\co{There are other proposals for probing the bulk topological phase transition instead of the edge states. Our method is new and all-electrical, allows to extract other parameters, and can be implemented in ongoing experiments.}
Our method is based on probing the bulk topological phase transition in Majorana devices, instead of the Majoranas themselves.
Several other works propose different methods to probe the bulk instead of the edge states in one-dimensional topological superconductors.
Quantized thermal conductance and electrical shot noise measurements are predicted signatures of a bulk topological phase transition \cite{akhmerov2011}, and here we present a different route based on straightforward electrical conduction measurements in already available experimental systems.
Further work predicts bulk signatures of a topological phase transition in the difference between the local Andreev conductances at each end of the proximitised region \cite{fregoso2013}, or in the spin projection of bulk bands along the magnetic field direction \cite{szumniak2017}.
In addition to probing the bulk topological phase transition, our proposed method allows to probe a number of relevant physical parameters, and can be implemented in ongoing experiments, providing a novel technique to use in the hunt for Majoranas.

\co{Descriptions of sections.}
This paper is organized as follows. In Sec.~\ref{sec:model}, we give an overview of our model and discuss the relevant energy and length scales.
In Sec.~\ref{sec:transport}, we study how nonlocal conductance measures superconductor characteristics.
We investigate the effects of a Zeeman field in homogeneous and inhomogeneous systems in Sec.~\ref{sec:rectifier}.
In Sec.~\ref{sec:cooperpairsplitter} we consider the possible application of the proximitised system as a Cooper pair splitter.
We finish with a summary and discussion of our results in Sec.~\ref{sec:summary}.

\section{Model and physical picture}
\label{sec:model}

\begin{figure}[!tbh]
\includegraphics[width=0.97\columnwidth]{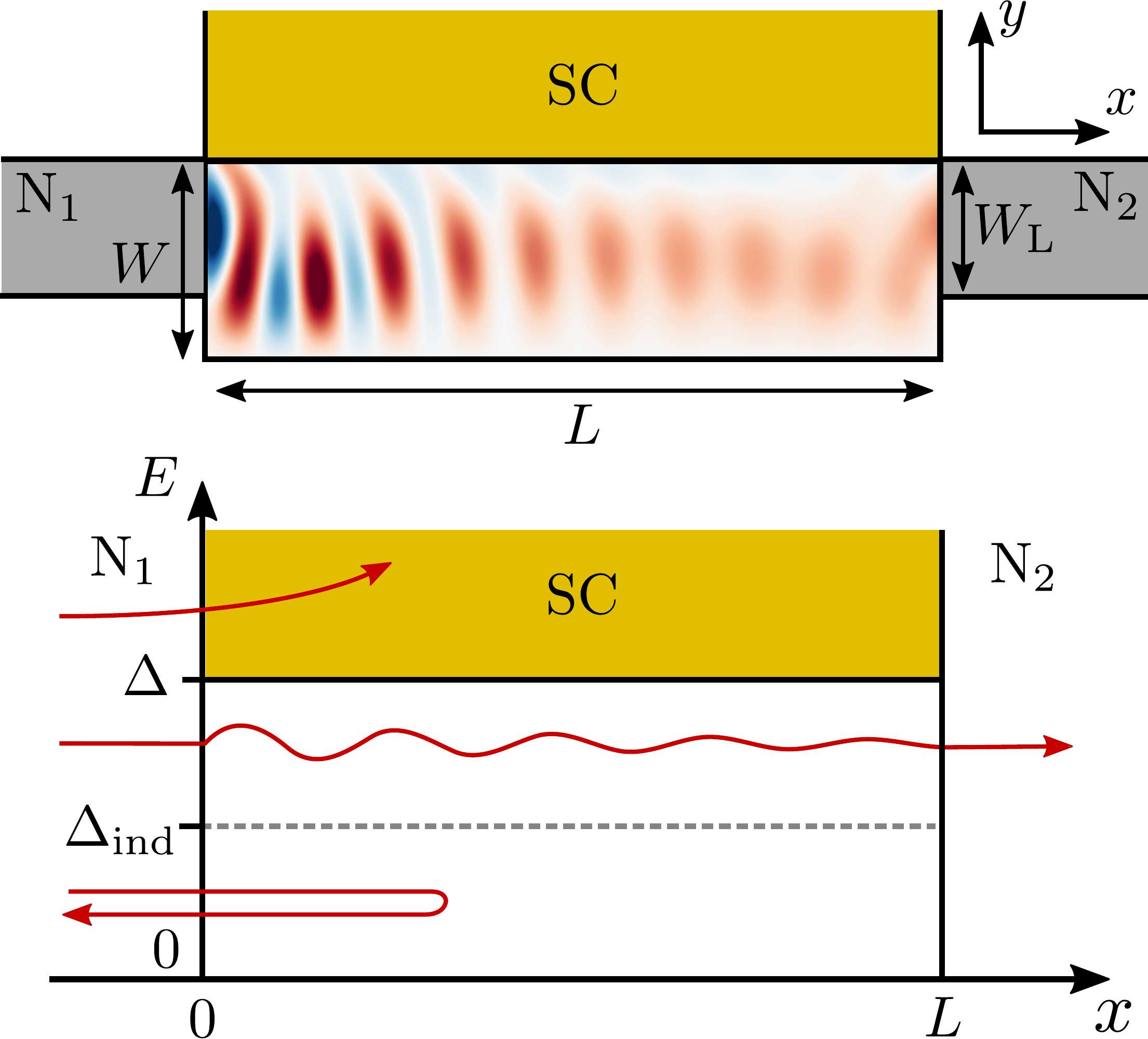}
\caption{Top: schematic drawing of the device.
A central region of length $L$ and width $W$ is connected from the sides to two normal leads N$_1$ (left) and N$_2$ (right) of width $W_\text{L}$, and from the top to one superconducting lead (SC) of width $L$.
Superimposed is an example of the central region charge density, which oscillates between positive (red) and negative (blue).
Bottom: illustrations of possible scattering processes in energy space in the limit $L \gg \xi$.
A quasiparticle, with energy below the induced gap, $|E| < \Delta_\text{ind}$, is reflected back into the source lead.
A quasiparticle at $\Delta_\text{ind} < |E| < \Delta$ is transmitted to the right lead, either as an electron (normal transmission) or as a hole (crossed Andreev reflection).
At energies exceeding the bulk gap $|E| > \Delta$, the superconducting lead absorbs incoming quasiparticles.}
\label{fig:setup}
\end{figure}

\co{The geometry consists of a proximitised normal material attached to two spatially separated normal leads.}
We consider a three terminal device sketched in Fig.~\ref{fig:setup}, with a normal central region of lateral length $L$ and width $W$ separating two normal leads of width $W_\text{L}$.
The device has a grounded superconducting lead of width $L$ attached to the central region perpendicularly to the other two leads.
This geometry models the proximity effect of a lateral superconductor on a slab of normal material, with normal leads probing the transport properties, and is therefore relevant both for heterostructures based on nanowires and quantum wells.

\co{The Hamiltonian consists of a kinetic term, chemical potential, spin-orbit interaction, superconducting pairing and Zeeman field.}
We model the hybrid system using the Bogoliubov-de Gennes Hamiltonian.
For a semiconductor electron band with effective mass $m^*$ and Rashba spin-orbit interaction (SOI) with strength $\alpha$, it reads
\begin{equation} \begin{split}
H = &\left( \frac{p_x^2 + p_y^2}{2m^*} - \mu\right)\tau_z  + \Delta (y) \tau_x + \\
& \frac{\alpha}{\hbar}\left( p_y\sigma_x - p_x \sigma_y \right)\tau_z + E_\text{Z} (y) \sigma_x, \label{eq:ham}
\end{split}\end{equation}
with $p_{x,y} = -i \hbar \partial_{x,y}$, $\mu$ the equilibrium chemical potential and $E_\text{Z}$ the Zeeman energy due to an in-plane magnetic field parallel to the interface between the central region and the superconductor.
We assume a constant $s$-wave pairing potential that is nonzero only in the superconductor, $\Delta (y) = \Delta \theta (y-W)$ with $\theta(y)$ a step function, and choose $\Delta$ to be real since only one superconductor is present.
We neglect the $g$-factor in the superconductor since it is much smaller than in the adjacent semiconductor, such that $E_\text{Z} (y) = E_\text{Z} \theta(W-y)$, and our conclusions are not affected by this choice.
The Pauli matrices $\sigma_i$ and $\tau_i$ act in spin and particle-hole space, respectively.
The Hamiltonian acts on the spinor $\Psi = \left( \psi_{e\uparrow}, \psi_{e\downarrow}, \psi_{h\downarrow}, -\psi_{h\uparrow} \right)$, which represent the electron (e) or hole (h) components of spin up ($\uparrow$) or down ($\downarrow$).

\co{The superconductor induces a gap in the proximitised system, the size of which is determined by the bulk gap and Thouless energy.}
The superconductor induces an energy gap $\Delta_{\mathrm{ind}}$ in the heterostructure.
If $L \gg W$, the larger of two energy scales, namely the bulk gap $\Delta$ and the Thouless energy $E_\text{Th}$, determines the magnitude of $\Delta_{\mathrm{ind}}$, with $E_\text{Th}$ at low $\mu$ given by
\begin{equation} E_\text{Th} = \gamma \delta, \hspace{0.2 cm} \delta = \frac{\hbar^2 \pi^2}{2m^* (2W)^2}, \end{equation}
where $\gamma$ is the transparency of the interface with the superconductor and $\delta$ the level spacing.
Our emphasis is on short and intermediate junctions, for which $E_\text{Th} \gg \Delta$ and $E_\text{Th} \lesssim \Delta$, respectively, such that $\Delta_{\mathrm{ind}} \lesssim \Delta$.
A brief review of normal-superconductor junctions in different limits and the relevant length and energy scales is given in Appendix~\ref{app:junctions}.
We keep $\mu$ constant in the entire system, but assume an anisotropic mass\cite{sticlet2016} in the superconductor with a component parallel to the interface $m_\parallel \rightarrow \infty$.
This approximation results in a transparent interface $\gamma=1$ at normal incidence and at $E_\text{Z}=0$, and is motivated by recent advances in the fabrication of proximitised systems with a high-quality superconductor-semiconductor interface \cite{kjaergaard2016, zhang2017}.

\co{Conductance is calculated using a scattering matrix formalism.}
We compute differential conductance using the scattering formalism.
The scattering matrix relating all incident and outgoing modes in the normal leads of Fig. \ref{fig:setup} is
\begin{equation}
S = \begin{bmatrix}
S_{11} & S_{12} \\
S_{21} & S_{22}
\end{bmatrix}, \hspace{0.2 cm}
S_{ij} = \begin{bmatrix}
S_{ij}^{ee} & S_{ij}^{eh} \\
S_{ij}^{he} & S_{ij}^{hh}
\end{bmatrix}. \label{eq:S}
\end{equation}
Here, the $S_{ij}^{\alpha \beta}$ is the block of scattering amplitudes of incident particles of type $\beta$ in lead $j$ to particles of type $\alpha$ in lead $i$.
Since quasiparticles may enter the superconducting lead for $|E|> \Delta$, the scattering matrix \eqref{eq:S} is unitary only if $|E|< \Delta$.
The zero-temperature differential conductance matrix equals \cite{blonder1982, anantram1996}
\begin{equation}
G_{ij} (E)\equiv \frac{\partial I_i}{\partial V_j} = \frac{e^2}{h} \left(
T_{ij}^{ee}- T_{ij}^{he} - \delta_{ij} N_i^e\right),
\label{eq:cond_0T}
\end{equation}
with $I_i$ the current entering terminal $i$ from the scattering region and $V_j$ the voltage applied to the terminal $j$, and  $N_j^e$ the number of electron modes at energy $E$ in terminal $j$, and finally the energy-dependent transmissions are
\begin{equation} T_{ij}^{\alpha \beta} = \mathrm{Tr}\left( \left[ S_{ij}^{\alpha \beta}(E)\right]^\dagger S_{ij}^{\alpha \beta} \right). \end{equation}
The blocks of the conductance matrix involving the superconducting terminal are fixed by the condition that the sum of each row and column of the conductance matrix has to vanish.
The finite temperature conductance is a convolution of zero-temperature conductance with a derivative of the Fermi distribution function $f(E) = (1+\exp{(E/k_BT)})^{-1}$:
\begin{equation}
G_{ij} (eV_j, T) = -\int_{-\infty}^{\infty} \mathrm{d}E \frac{\mathrm{d} f(E-eV_j, T)}{\mathrm{d}E} G_{ij}(E).
\label{eq:cond} \end{equation}

\co{We use Kwant to compute the scattering matrix and $\xi$, using material parameters for an InAs 2DEG with epixatial Al.}
We discretize the Hamiltonian \eqref{eq:ham} on a square lattice, and use Kwant \cite{groth2014} to numerically obtain the scattering matrix~of Eq.~\eqref{eq:S}, see the supplementary material for source code \cite{suppl_code}.
The resulting data is available in Ref.~\onlinecite{suppl_data}.
We obtain $\xi$ numerically by performing an eigendecomposition of the translation operator in the $x$-direction for a translationally invariant system and computing the decay length of the slowest decaying mode at $E=0$ \cite{nijholt2016, sticlet2016}.
We use the material parameters\footnote{All parameters are provided per figure in a text file as supplementary material.} $m^* = 0.023 m_\text{e}$, $\alpha = 28$ meVnm, and unless otherwise specified $\Delta = 0.2$ meV, typical for an InAs two-dimensional electron gas with an epitaxial Al layer \cite{kjaergaard2016}.
All transport calculations are done using $T = 30$ mK unless stated otherwise.

\section{Nonlocal conductance as a measure of superconductor properties}
\label{sec:transport}

\begin{figure}[!tbh]
\includegraphics[width=0.97\columnwidth]{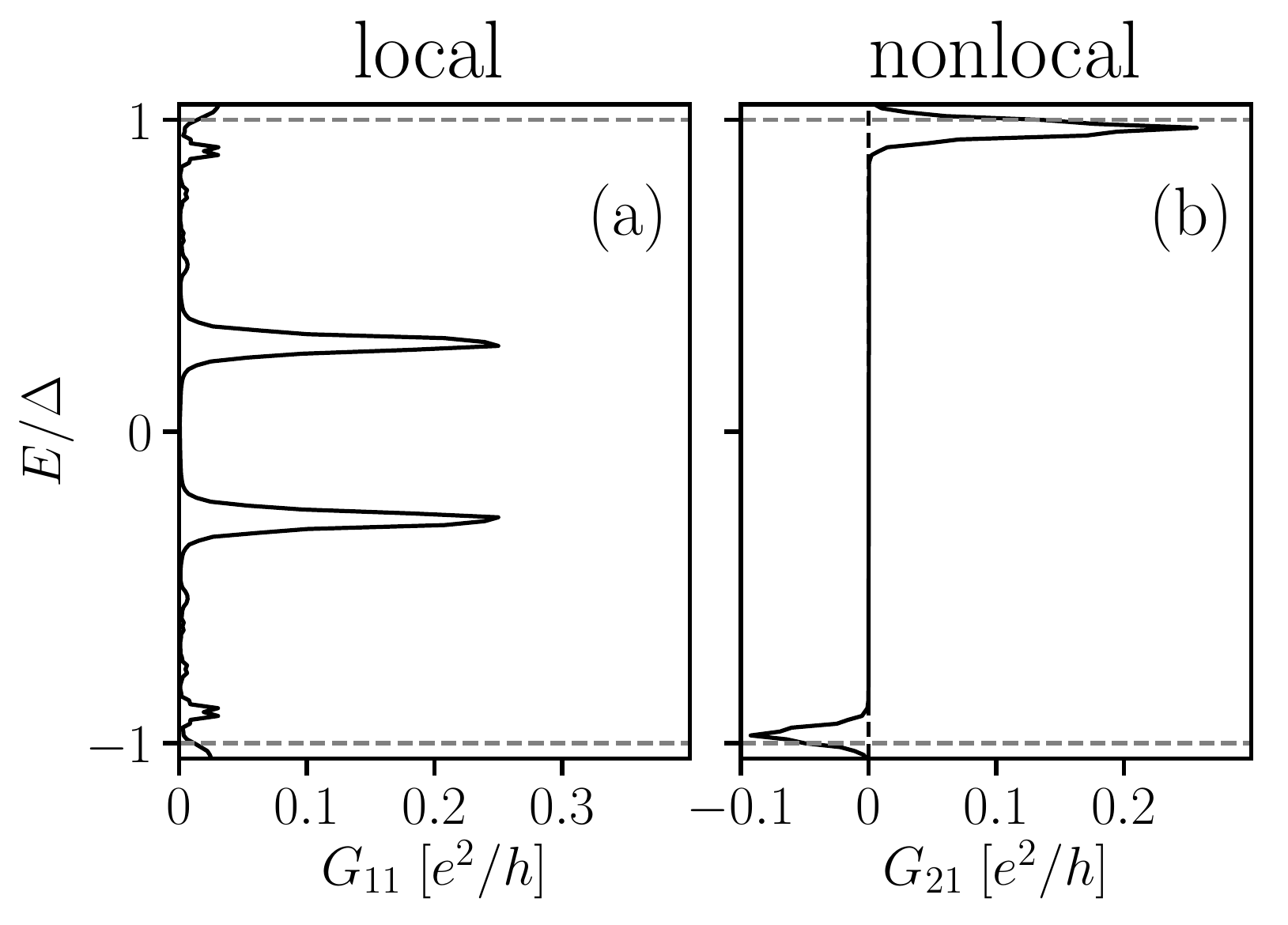}
\caption{Examples of (a) the local conductance in the tunnelling regime and (b) the nonlocal conductance in the open regime, of an inhomogeneous proximitised system with broken time-reversal symmetry.
Localized low-energy states are present near the junctions with the normal leads.
These manifest as peaks in the tunnelling conductance, indicating $\Delta_{\mathrm{ind}} \ll \Delta$.
However, $\Delta_{\mathrm{ind}} \approx \Delta$ still in the bulk of the proximitised system, with $\Delta_{\mathrm{ind}}$ matching the energy at which the nonlocal conductance peaks.}
\label{fig:inhomo_line_cuts}
\end{figure}

\co{Local conductance tunnel spectroscopy is not a reliable probe of the bulk properties of induced superconductivity, since it only probes the density of states close to the probe. On the other hand, nonlocal conductance probes the bulk properties.}
In the tunnelling regime, the local conductance in a normal lead probes the density of states in the proximitised region, which is commonly used to measure the induced gap in experiment.
However, such a measurement only probes the region near the tunnel probe, but fails to give information about the density of states in the bulk of the proximitised region.
The tunnelling conductance is thus not a reliable probe of the entire proximitised region if the density of states varies spatially over the proximitised region, for example due to an inhomogeneous geometry.
As an illustration, Fig.~\ref{fig:inhomo_line_cuts} compares the local conductance $G_{11}$ in the tunnelling limit to the nonlocal conductance $G_{21}$ in the open regime for a proximitised system that is inhomogeneous and in a magnetic field.
Inhomogeneous systems are further treated in Sec.~\ref{sec:rectifier}.
The combination of an inhomogeneous system and broken time-reversal symmetry creates low-energy states localized near the junctions with the normal leads, which appear as peaks in the tunnelling conductance.
However, away from the junctions with the normal leads, the proximitised system remains close to fully gapped, the induced gap matching the energies at which the nonlocal conductance becomes finite in Fig.~\ref{fig:inhomo_line_cuts}(b).
Therefore, the nonlocal conductance is better than the local tunnelling conductance as a probe for the induced gap in the bulk of the proximitised region.
In the following, we describe three ways in which the nonlocal conductance probes induced superconductivity.

\begin{figure}[!tbh]
\includegraphics[width=0.97\columnwidth]{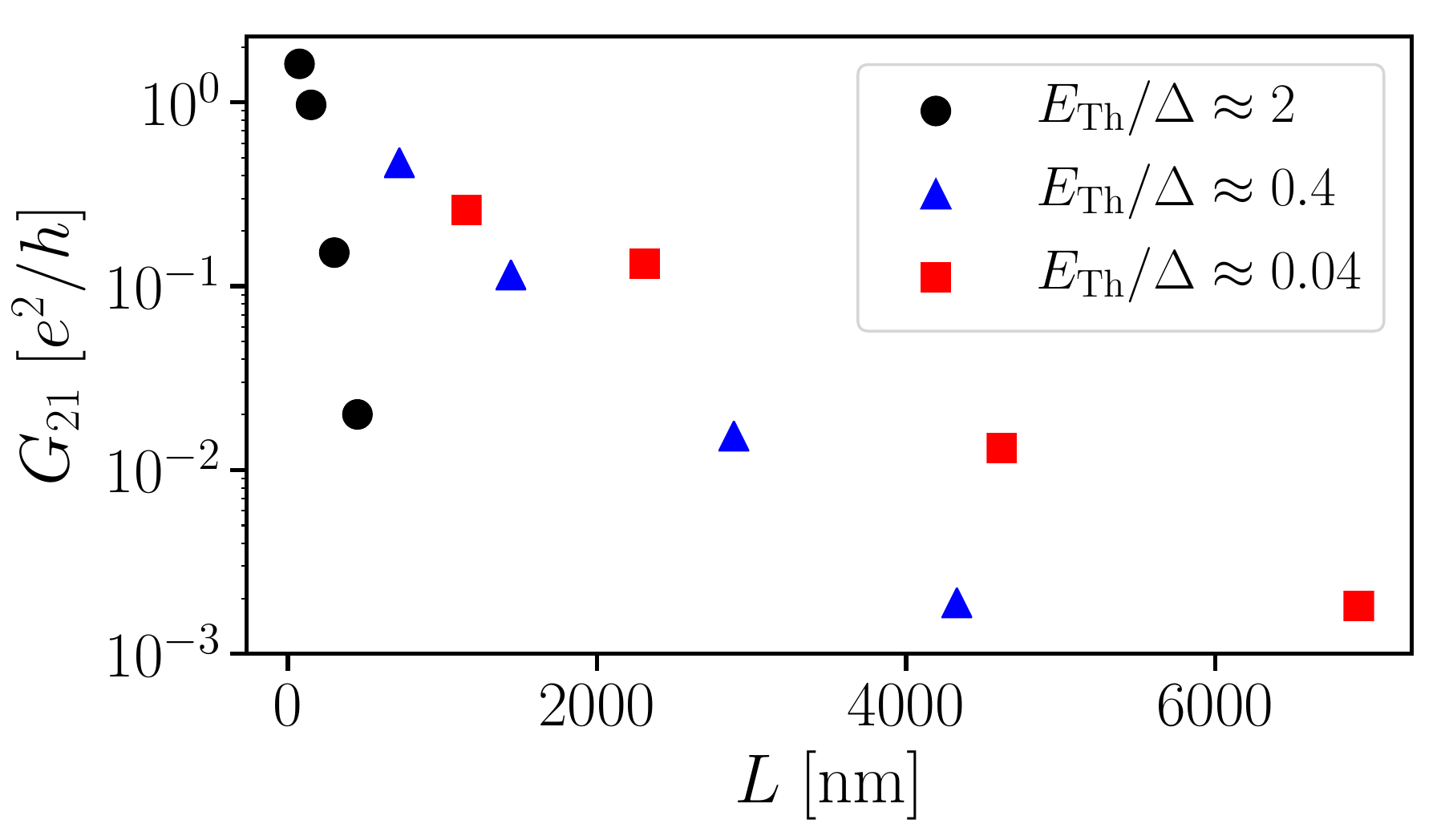}
\caption{Suppression of the nonlocal conductance $G_{21}$ at zero bias $E=0$ as a function of length for different ratios $E_\text{Th}/\Delta$. For decreasing ratio $E_\text{Th}/\Delta$, the induced coherence length $\xi$ increases. This is reflected in the larger absolute length over which the nonlocal conductance is suppressed. Data points are taken from the $E=0$ values of the nonlocal conductance presented in Fig.~\ref{fig:disp_nonloc_no_disorder}.}
\label{fig:nonloc_vs_length_zero_bias}
\end{figure}

\co{Nonlocal conductance is suppressed at $E=0$ for $L > \xi$, hence it measures the ratio between $L$ and $\xi$.}
First of all, the nonlocal conductance measures the induced decay length $\xi$ in the bulk of the proximitised region between the two normal leads.
To understand this, consider a nonlocal process at a subgap energy $\left| E \right| < \Delta_{\mathrm{ind}}$.
An electron injected from a normal lead must propagate as an evanescent wave $\propto e^{-x/\xi + ikx}$ through the gapped central region to the second normal lead, with $\xi$ the decay length.
Accordingly, as shown in Fig.~\ref{fig:nonloc_vs_length_zero_bias}, increasing $L$ suppresses the nonlocal conductance at $E = 0$ exponentially \cite{deutscher2000, melin2002}.
Therefore, the suppression of the nonlocal conductance with increasing length $L$ at $E=0$ is a measure of the induced decay length $\xi$.

\begin{figure}[!tbh]
\includegraphics[width=0.97\columnwidth]{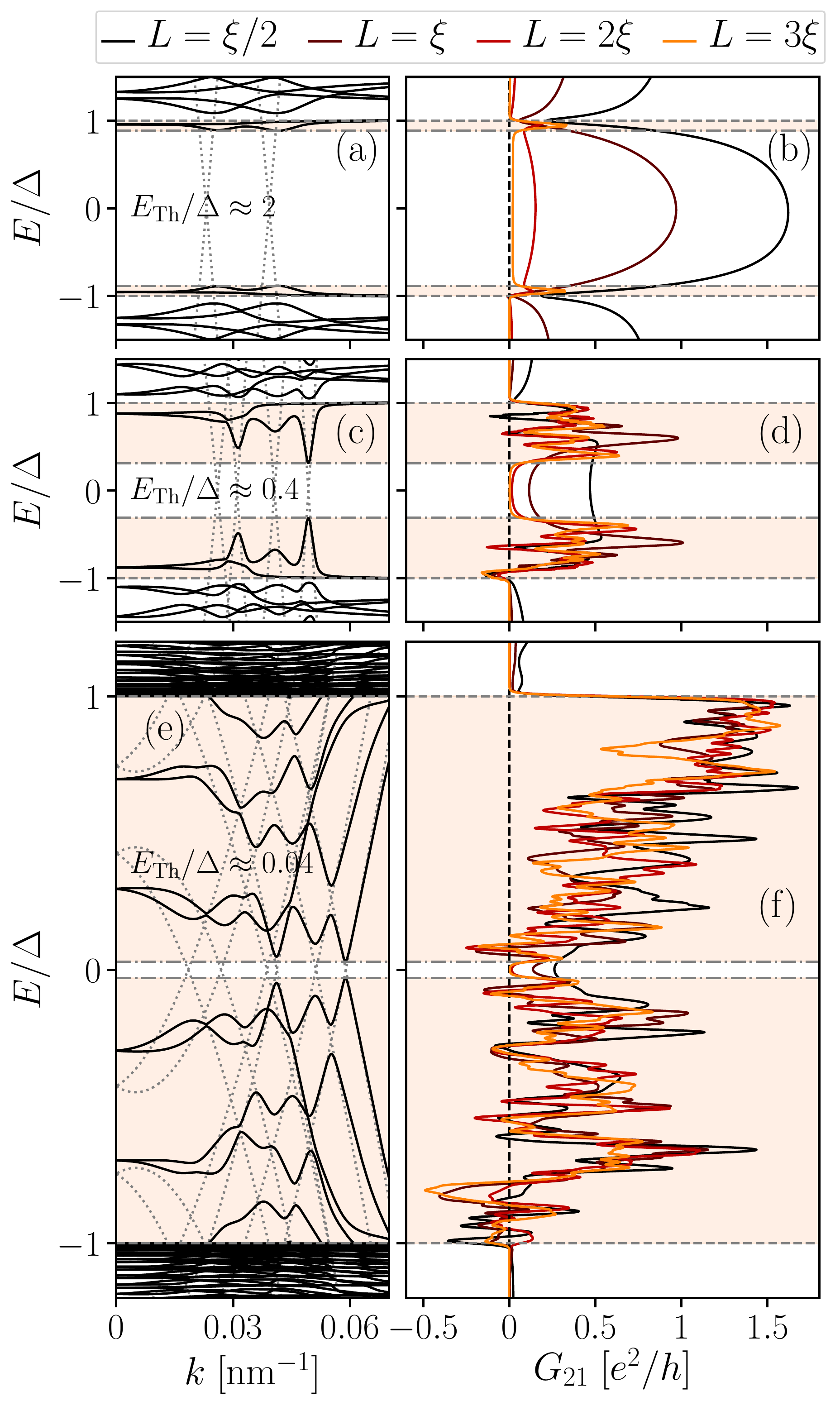}
\caption{(a, c, e) Dispersions of proximitised systems that are translationally invariant along $x$, and (b, d, f) nonlocal conductance $G_{21}(E)$ of corresponding junctions of finite length.
The latter is shown as the separation $L$ between the two normal leads is varied, with brightening colors from black to orange denoting $L = \xi/2, \ \xi, \ 2\xi$ and $3\xi$, respectively.
The ratio $E_{\mathrm{Th}}/\Delta$ becomes smaller from top to bottom, such that $\Delta_{\mathrm{ind}}$ shrinks (dash-dotted lines).
For $L \gg \xi$, the nonlocal conductance is suppressed if $\left| E \right| < \Delta_{\mathrm{ind}}$, and finite only for $\Delta_{\mathrm{ind}}<\left| E \right| < \Delta$ (colored region).
The solid lines in the dispersion relations show the dispersion of the normal-superconductor system, while the dotted lines show the electron and hole dispersion of the normal channel only, with the superconductor removed.
We have $W = 100$ nm in (a) and (b), $W = 200$ nm in (c), (d), (e) and (f), and $W_L = 100$ nm always in the right column.
$\mu = 3$ meV, $\Delta = 0.2$ meV and $T = 30$ mK in the top and middle rows, but $\mu =4.2$ meV, $\Delta = 2$ meV and $T = 100$ mK in the bottom row. Dispersions are even in $k$.}
\label{fig:disp_nonloc_no_disorder}
\end{figure}

\co{For $L \gtrsim \xi$, nonlocal conductance is zero for $E>\Delta$, hence it measures the bulk superconducting gap}
Furthermore, the nonlocal conductance measures the bulk gap $\Delta$ of the superconductor.
Increasing $L$ also suppresses the nonlocal conductance $G_{21}$ for $|E| > \Delta$, as the right column of Fig.~\ref{fig:disp_nonloc_no_disorder} shows.
For energies above the bulk superconducting gap $\Delta$, the superconductor increasingly absorbs quasiparticles when the length is increased, and suppresses the nonlocal conductance to zero when $L \gg \xi$.
Hence, the energy above which nonlocal conductance is suppressed at large lengths is a measure of $\Delta$.

\co{Since, for $L \gtrsim \xi$, nonlocal conductance is only nonzero for $\Delta_\text{ind} < E < \Delta$, it also measures the induced gap.}
In addition, the nonlocal conductance measures the induced superconducting gap $\Delta_\text{ind}$.
When $L \gtrsim \xi$, the nonlocal conductance is suppressed at $E = 0$ but grows in a convex shape with $E$ and peaks around $\left| E \right| \approx \Delta_{\mathrm{ind}}$, as shown in the right column of Fig.~\ref{fig:disp_nonloc_no_disorder}.
This is due to a divergence in $\xi$, since the system is no longer gapped.
To illustrate the correspondence between the nonlocal conductance and $\Delta_{\mathrm{ind}}$, the left column of Fig.~\ref{fig:disp_nonloc_no_disorder} shows the dispersions of the corresponding proximitised systems that have the normal leads removed and are translationally invariant along the $x$ direction, such that $k = p_x/\hbar$ is conserved.
Because the system is not gapped for $\left|E \right| > \Delta_\text{ind}$, $G_{21}$ is generally nonzero at these energies.
Note that aside from occasional dips to negative $G_{21}$, direct electron transfer dominates the nonlocal response (we investigate this in more detail in Sec.~\ref{sec:cooperpairsplitter}).

\begin{figure}[!tbh]
\includegraphics[width=0.97\columnwidth]{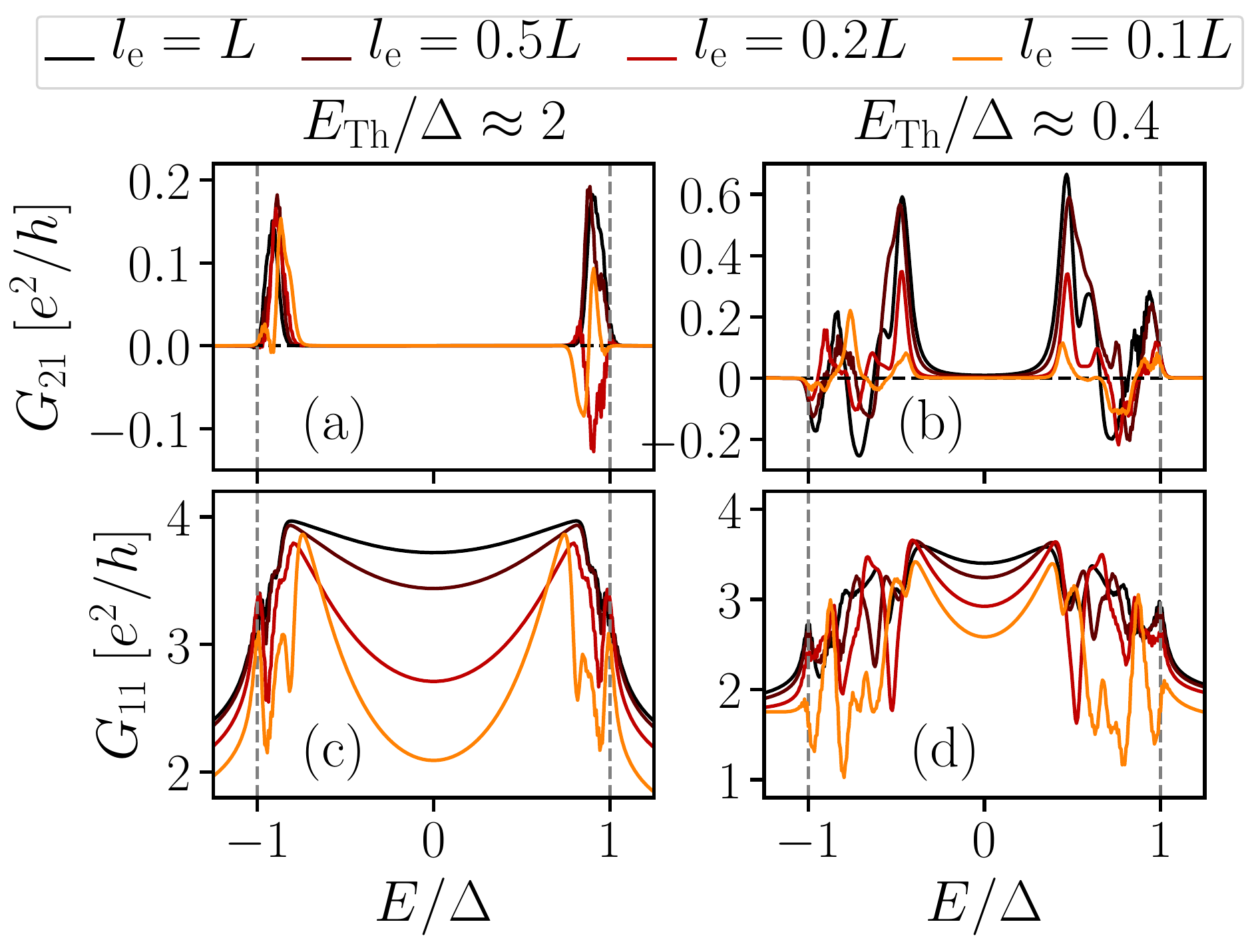}
\caption{Nonlocal (a, b) and local conductance (c, d) $G_{21}$ and $G_{11}$ of short (left column) and intermediate (right column) junctions with $L \gtrsim \xi$, to decouple the two normal leads at $\left| E \right| < \Delta_{\mathrm{ind}}$.
The mean free path varies between curves, with brightening colors from black to light orange denoting $l_\text{e} = L, \ L/2,  \ L/5$ and $L/10$ respectively.
Even in the presence of disorder, signatures of $\Delta_{\mathrm{ind}}$ and $\Delta$ are visible in the nonlocal conductance.
The local conductance is Andreev enchanced at subgap energies, but normal reflection becomes more prominent with increasing disorder.
We have $W = 100$ nm and $L = 8 \xi$ in (a) and (c), $W=200$ nm and $L = 2\xi$ in (b) and (d), and $W_L=100$ nm always, with $\mu = 3$ meV.}
\label{fig:loc_nonloc_disorder}
\end{figure}

\co{All of the above still holds in the presence of disorder, such that the nonlocal conductance remains a reliable probe of induced superconductivity.}
The presence of finite nonlocal conductance in the energy range  $\Delta_\text{ind} < \left|E \right| < \Delta$ depends only on density of states of the proximitised system, and therefore still holds in the presence of disorder.
In Fig.~\ref{fig:loc_nonloc_disorder}, we show the effects of disorder on the transport signatures of $\Delta$ and $\Delta_{\mathrm{ind}}$ for short and intermediate junctions when $L \gtrsim \xi$.
We include onsite disorder in the central region, and vary the elastic mean free path $l_\text{e}$ from $l_\text{e} = L$ to $l_\text{e} = 0.1L$ \cite{ando1991}.
Even in the presence of disorder, all of the aforementioned qualities are still apparent in the nonlocal conductance (a) and (b), namely suppression for $\left| E \right| < \Delta_{\mathrm{ind}}$, a finite signal for $\Delta_{\mathrm{ind}} < \left| E \right| < \Delta$ and vanishing conductance for $\left|E \right| > \Delta$.
Therefore, the nonlocal conductance remains a reliable probe of induced superconductivity even in the presence of disorder.

\co{The open-regime local conductance is Andreev enchanced and smooth at subgap energies, and the bulk and induced gaps are visible in the local conductance.
However, this is only because there are no extended potential inhomogeneities.}
Lastly, in the absence of extended potential inhomogeneities, $\Delta$ and $\Delta_{\mathrm{ind}}$ may also be inferred from the local conductance $G_{11}$ in the open regime.
As Figs.~\ref{fig:loc_nonloc_disorder}(c) and (d) show, $G_{11} \lesssim 4e^2/h$ in the ballistic case $l_\text{e} = L$ for $\left| E \right| < \Delta_{\mathrm{ind}}$, which indicates that Andreev reflection is the dominant local process.
This is the expected behavior for a normal-superconductor junction with high interface transparency \cite{blonder1982, kjaergaard2016}, and is consistent with our results.
Reducing the mean free path makes normal reflection more likely and hence lowers $G_{11}$, similar to an ideal normal-superconductor junction with a reduced interface transparency.
Here, comparing $G_{11}$ and $G_{21}$ shows that $\Delta_{\mathrm{ind}}$ and $\Delta$ may also be inferred from the local conductance, because it changes smoothly with bias only outside the interval $\Delta_{\mathrm{ind}}< \left| E \right| < \Delta$.
However, the signatures are clearer in $G_{21}$, where it is a transition between finite and vanishing conductance that indicates the gaps.
Furthermore, the induced gap observed in the local and nonlocal conductances coincide here only due to the absence of extended potential inhomogeneities.
For the case of an inhomogeneous geometry as in Fig.~\ref{fig:inhomo_line_cuts}, only the nonlocal conductance correctly measures $\Delta_{\mathrm{ind}}$ in the bulk of the proximitised system.

\section{Andreev rectifier at the topological phase transition}
\label{sec:rectifier}

\subsection{Andreev rectification as a measure of the topological phase}
\co{The proximitised system acts as a rectifier of the applied voltage at the topological phase transition.}

\begin{figure}[!tbh]
\includegraphics[width=0.97\columnwidth]{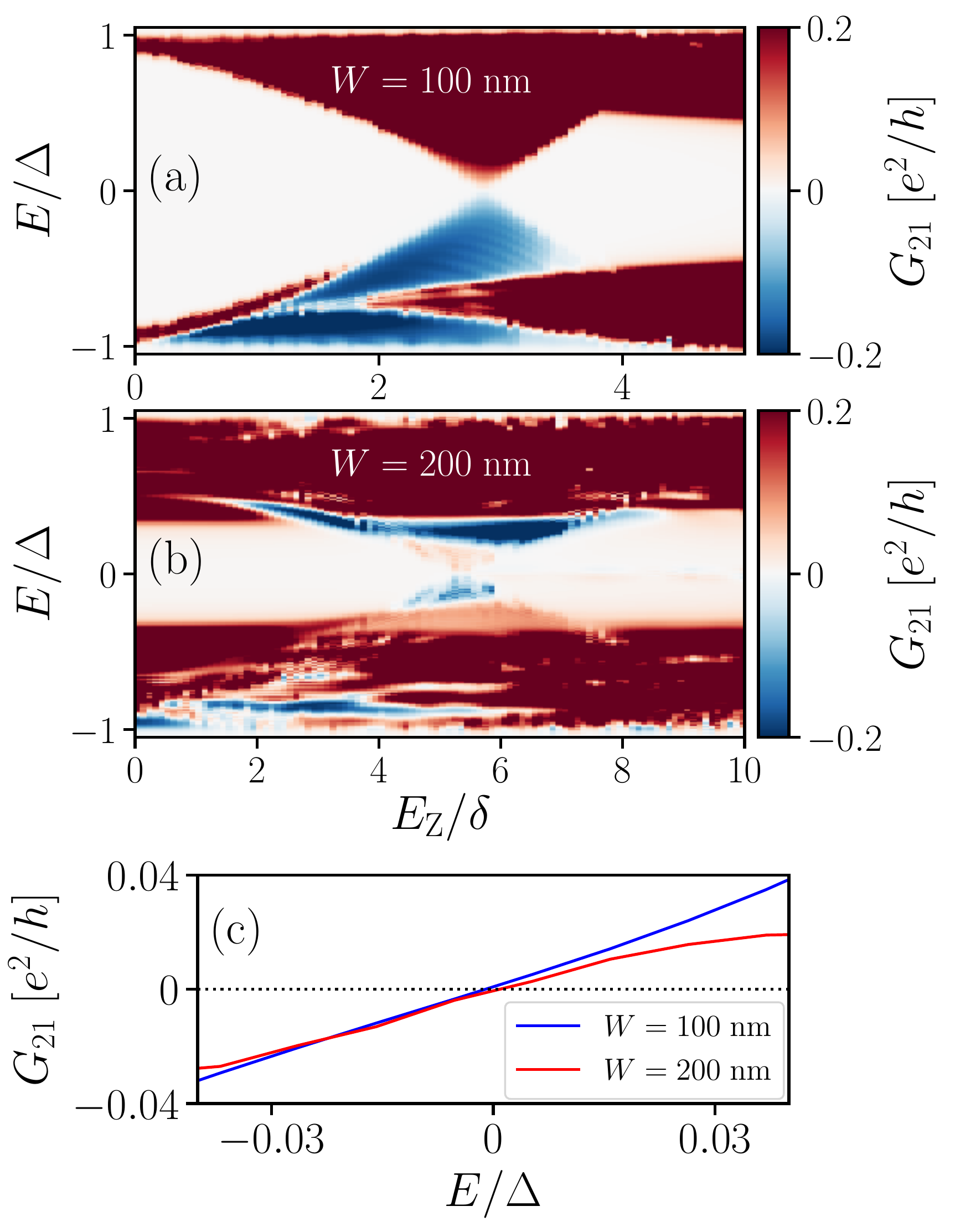}
\caption{(a, b) Nonlocal conductance $G_{21}$ in the single-mode regime as a function of $E$ and $E_\text{Z}$ in the absence of disorder.
We have $W = 100$ and $200$~nm in (a) and (b), respectively.
The Zeeman field closes the induced gap and the system undergoes a topological phase transition.
At the transition, $G_{21}$ vanishes and changes sign as a function of bias.
There are prominent regions where the nonlocal conductance is negative, i.e.\ where CAR dominates.
The color scale is saturated for clarity.
(c) Line cuts of $G_{21}$ as a function of bias at the topological phase transition, taken at $E_\text{Z} \approx 2.9\delta$ for $W = 100$~nm and $E_\text{Z} \approx 5.4 \delta$ for $W = 200$~nm, showing that the nonlocal conductance is an approximately odd function of bias.}
\label{fig:loc_nonloc_no_dis_color}
\end{figure}

In order to study nonlocal conductance at the topological phase transition, we apply an in-plane Zeeman field along the $x$-direction of the proximitised system.
Figure \ref{fig:loc_nonloc_no_dis_color} shows the nonlocal conductance $G_{21}$ as a function of bias $E$ and Zeeman energy $E_\text{Z}$, for short and intermediate junctions in (a) and (b) with $L = 10\xi$ and $L = 3\xi$, respectively, such that the two normal leads are well decoupled, and the nonlocal conductance is exponentially suppressed at subgap energies.
Increasing the magnetic field closes the induced gap and the system is driven into a topological phase.
The line cuts of Fig.~\ref{fig:loc_nonloc_no_dis_color}(c), taken at the critical magnetic field $E_\mathrm{Z} =E_\mathrm{Z}^c$, show that at the topological phase transition the nonlocal conductance is a linear function of energy, $G_{21}(E) \propto E$ around $E=0$.
At the topological phase transition, the current $I \propto V^2$ with $V$ the voltage bias, and the system functions as a current rectifier due to crossed Andreev reflection.

\co{The linear behaviour of the conductance is due to topology and symmetry, and makes the junction a rectifier at the topological phase transition.}
This Andreev rectifier manifests due to the topology and symmetry of the proximitised system.
The system only has particle-hole symmetry and is therefore in class $D$ \cite{hell2016, pientka2016}.
Expanding $G_{21}(E, E_\mathrm{Z}) = c_0(E_\mathrm{Z}) + c_1(E_\mathrm{Z}) E + O(E^2)$ around $E = 0$, the exponential suppression of $G_{21}$ at subgap energies means that the coefficients $c_0$ and $c_1$ are exponentially suppressed at magnetic fields before the topological phase transition.
In class $D$ systems, if $G_{21}$ is exponentially suppressed at subgap energies, it is guaranteed to remain exponentially suppressed across the topological phase transition \cite{akhmerov2011, wieder2014}.
At the critical magnetic field $E_\mathrm{Z} = E_\mathrm{Z}^c$, $G_{21}(E=0, E_\mathrm{Z}^c) = c_0(E_\mathrm{Z}^c)$ is therefore also exponentially suppressed.
However, the system is gapless at the topological phase transition, such that $G_{21}$ is generally finite at any nonzero $E$, and $c_{1}(E_\mathrm{Z}^c)$ thus not exponentially suppressed.
At the topological phase transition, we therefore have $G_{21} \propto E$ in the limit $E \rightarrow 0$, where higher order contributions are negligible.
Consequently, rectifying behavior in the nonlocal conductance is an indication of a topological phase transition.
This makes the nonlocal conductance not only a probe of the bulk properties of induced superconductivity as discussed in Sec.~\ref{sec:transport}, but also makes it selectively sensitive to topological phase transitions.

\begin{figure}[!tbh]
\includegraphics[width=0.97\columnwidth]{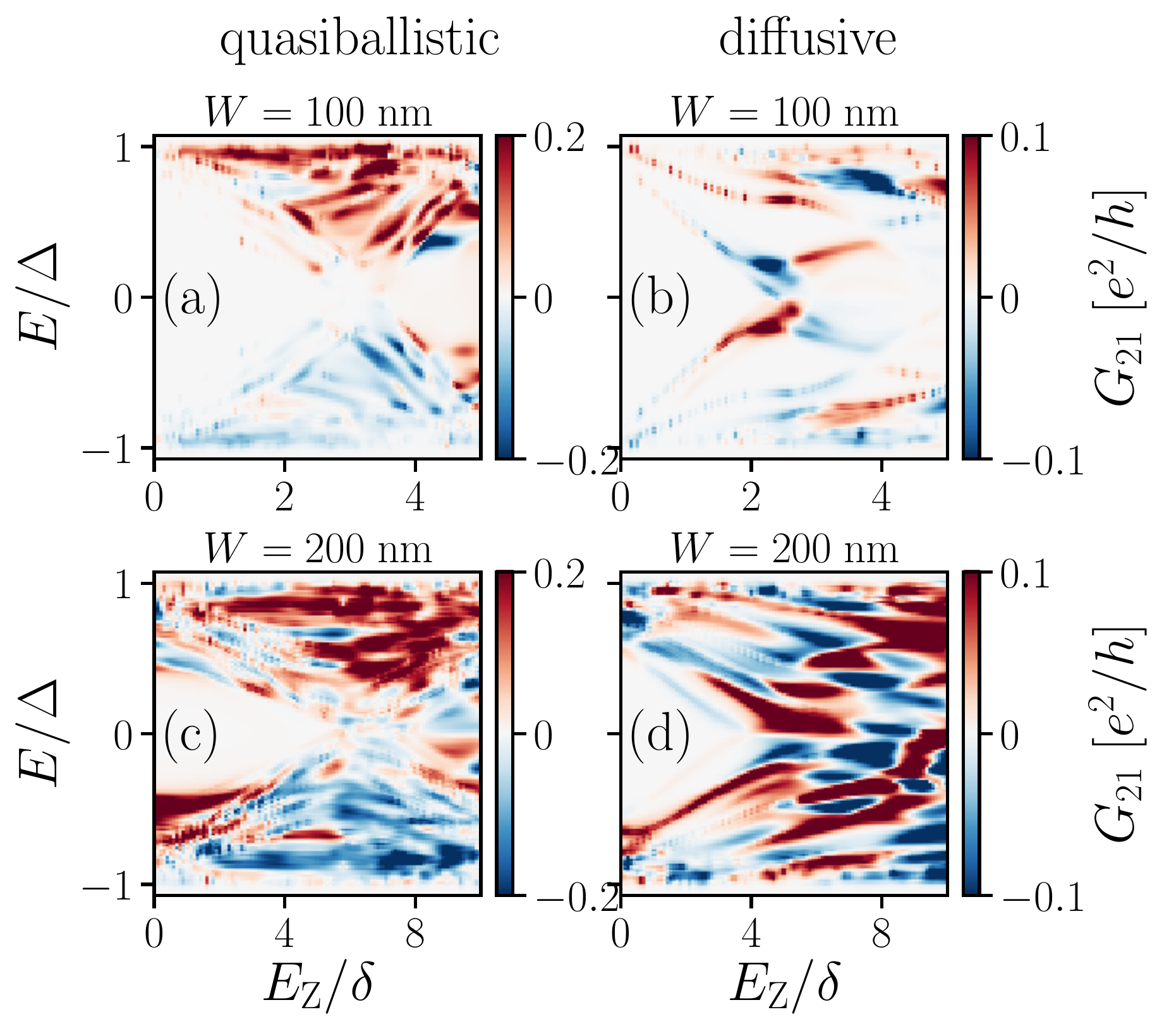}
\caption{The nonlocal conductance $G_{21}$ as a function of $E$ and $E_\text{Z}$ for a proximitised system that is quasiballisic (a, c with $\mu = 3$~meV) and diffusive (b, d with $\mu=16$~meV).
For the diffusive junction, the leads are gated into the single-mode regime using quantum point contacts at the junctions with the channel.
Top and bottom row present results for $W = 100$ and $200$~nm respectively.
For the quasiballistic junction, $L/\xi = 8$ and $2$ for $W = 100$ nm and $200$ nm respectively, and the mean free path is $l_\text{e} = 0.2L$ in each case.
In the diffusive system, we have $l_\text{e} = 0.2W$ and $L/\tilde{\xi} = 5$ and $2$ for the widths respectively, where $\tilde{\xi} = \sqrt{l_\text{e} \xi}$.
The color scale is saturated in both cases for clarity.}
\label{fig:nonloc_dis_color}
\end{figure}

\co{The Andreev rectifier is robust against disorder.}
The rectifying behavior $G_{21} \propto E$ at the topological phase transition in Fig.~\ref{fig:loc_nonloc_no_dis_color} is grounded in the symmetry classification of the channel.
As a result, we expect it to be robust to the presence of onsite disorder, so long as it does not alter the symmetry class.
Figure \ref{fig:nonloc_dis_color} shows $G_{21}$ as a function of $E$ and $E_\text{Z}$ for systems with the same widths as in Fig.~\ref{fig:loc_nonloc_no_dis_color}.
In the left column, parameters are chosen identical to those in Fig.~\ref{fig:loc_nonloc_no_dis_color}, with the addition of onsite disorder to give an elastic mean free path $l_\text{e} = 0.2 L$\cite{ando1991}, bringing the systems well into the quasiballistic regime.
In the right column of Fig.~\ref{fig:nonloc_dis_color}, we investigate $G_{21}$ when the central region is in the diffusive limit, with $l_\text{e} = 0.2 W$.
The widths are the same as in the quasiballistic (and clean) case, but $\mu$ is increased such that several modes are active.
We gate the leads into the single mode regime using quantum point contacts at the junctions with the scattering region.
In each case we pick $L \gtrsim \tilde{\xi}$, since in the diffusive limit $\tilde{\xi} = \sqrt{\xi l_\text{e}}$ governs the range of the coupling between the two normal terminals at subgap energies \cite{feinberg2003}.
In both quasiballistic and diffusive cases, $G_{21}$ remains an approximately odd function of $E$ around the gap closing, and the proximitised system therefore acts as a rectifier even in the presence of disorder.

\subsection{Distinguishing the topological phase transition in spatially inhomogeneous devices}

\co{Recent literature points out that local conductance measurements do not distinguish between trivial and topological zero-energy states in spatially inhomogeneous devices.}
Several works \cite{pikulin2012, Kells2012, mi2014, Prada2012, Moore2016, liu2017} discuss the emergence of zero-energy modes in the trivial phase of a hybrid semiconductor-superconductor device with an extended, spatially inhomogeneous potential.
Local conductance measurements do not distinguish between these modes and well-separated Majorana modes at the endpoints of the proximitised region in the topological phase, since both give rise to zero-bias conductance features.

\co{To study this issue, we consider the nonlocal conductance in systems with a potential inhomogeneity.}
To study this problem, we include an extended inhomogeneous potential
\begin{equation}
\phi(x, y) = V_0 \mathrm{exp}\left[-\frac{1}{2}\left(\frac{x - x_0}{d_x}\right)^2\right] \mathrm{exp}\left[ -\frac{1}{2}\left(\frac{y - y_0}{d_y}\right)^2\right],
\label{eq:Gausspot}
\end{equation}
in the setup shown in Fig.~\ref{fig:setup}, with $V_0$ the potential amplitude, $x_0$ and $y_0$ the coordinates of the potential center, and $d_x$ and $d_y$ parameters to control the smoothness in $x$- and $y$-direction, respectively.
We compare conductance for an amplitude $V_0 = -4.5$~mV to conductance in a homogeneous system $V_0 = 0$~V.
We calculate the local conductance in the tunneling regime, with tunnel barriers at both wire ends $x=0$ and $x=L$, and the nonlocal conductance in the open regime, with the system length fixed to $L = 8\xi$ and the width to $W = 100$~nm.
\begin{figure}[tb]
\includegraphics[width=0.97\columnwidth]{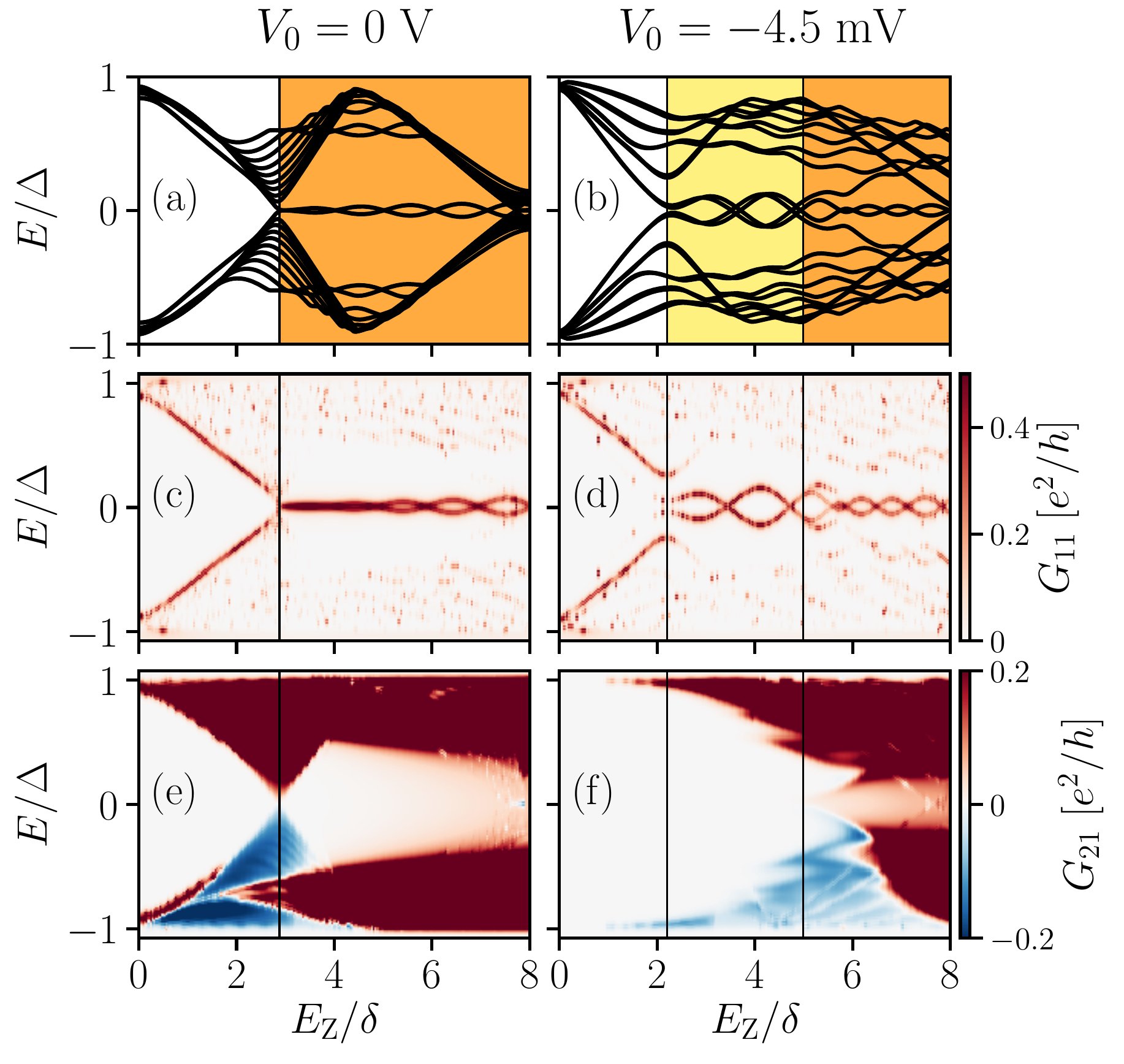}
\caption{Spectrum (a, b), local conductance $G_{11}$ (c, d) and nonlocal conductance $G_{21}$ (e, f) of a system without potential variations (left column) and a system with a long-range Gaussian potential of amplitude $V_0 = -4.5$~mV (right column).
The orange region in (a) and (b) denotes the topological phase, yellow the trivial phase with a state around zero energy.
$G_{11}$ is calculated in presence of two tunnel barriers at both wire ends, $G_{21}$ in the single mode regime.
The color scale is saturated for clarity.
For the potential inhomogeneity, we set $V_0 = -4.5$ meV, $x_0 = L/2$, $y_0 = W/2$, $d_x = L/5$ and $d_y = 2W/3$.}
\label{fig:broken_wire}
\end{figure}
\co{The spectrum of such system confirms that such potentials can lead to gap closings and zero-energy modes in the trivial regime.}
To confirm that such a spatially inhomogeneous system can indeed exhibit trivial zero-energy modes, we calculate the low-energy spectrum of our system when decoupled from the leads, forming a closed superconductor-semiconductor system.
The phase transition is computed from the absolute value of the determinant of the reflection matrix in the open system at $E=0$, with  $|\mathrm{det}(r)| = 1$ everywhere for $L \gg \xi$, except at the phase transition, where it drops to zero \cite{fulga2011}.
Fig.~\ref{fig:broken_wire}(a) shows the spectrum as a function of $E_\text{Z}$ in the homogeneous case ($V_0=0$), Fig.~\ref{fig:broken_wire}(b) for the inhomogeneous case ($V_0 = -4.5$~mV).
While in the first case the closing of the induced superconducting gap coincides with the topological phase transition, in the second case an extended topologically trivial region exists with states around zero energy (yellow region).

\co{Local conductance can not distinguish topological from non-topological gap closings.}
Comparing the local conductance with and without an inhomogeneous potential, we find that zero-energy modes appear regardless of whether they are topological or trivial. 
Panels (c) and (d) of Fig.~\ref{fig:broken_wire} show the local response as a function of bias and Zeeman energy when leads are connected to the central region via tunnel barriers.
Since the system is ballistic and long ($L \gg \xi$), the local conductance agrees well with the spectra presented in panels (a) and (b).
Accordingly, the local conductance in panel (d) for $V_0 = -4.5$~mV shows zero-energy modes in the topologically trivial regime.
Therefore, a gap closing and the emergence of zero-energy modes in the local conductance is not a sufficient sign of a topological phase transition.

\co{However, due to its rectifying behavior, nonlocal conductance is a more reliable measure of a topological phase transition.}
On the other hand, nonlocal conductance has a much clearer signature of the topological transition than the local conductance.
To demonstrate this, in panels (e) and (f) of Fig.~\ref{fig:broken_wire} we show the nonlocal conductance as a function of bias and Zeeman energy.
Both for the homogeneous and the inhomogeous case, the appearance of nonlocal conductance around $E=0$ coincides with the change of the topological invariant.
In other words, the appearance of finite nonlocal conductance around $E = 0$ implies a global closing of the induced gap.
Additionally, the nonlocal conductance shows rectifying behavior around $E=0$ at the gap closing.
These two features of the nonlocal conductance are strong evidence of a topological phase transition.
Therefore, due to its insensitivity to spatial inhomogeneities in the potential and the additional feature of Andreev rectification, nonlocal conductance is a more reliable measure of a topological phase transition.

\section{Cooper pair splitter}
\label{sec:cooperpairsplitter}
\begin{figure}[!tbh]
\includegraphics[width=0.97\columnwidth]{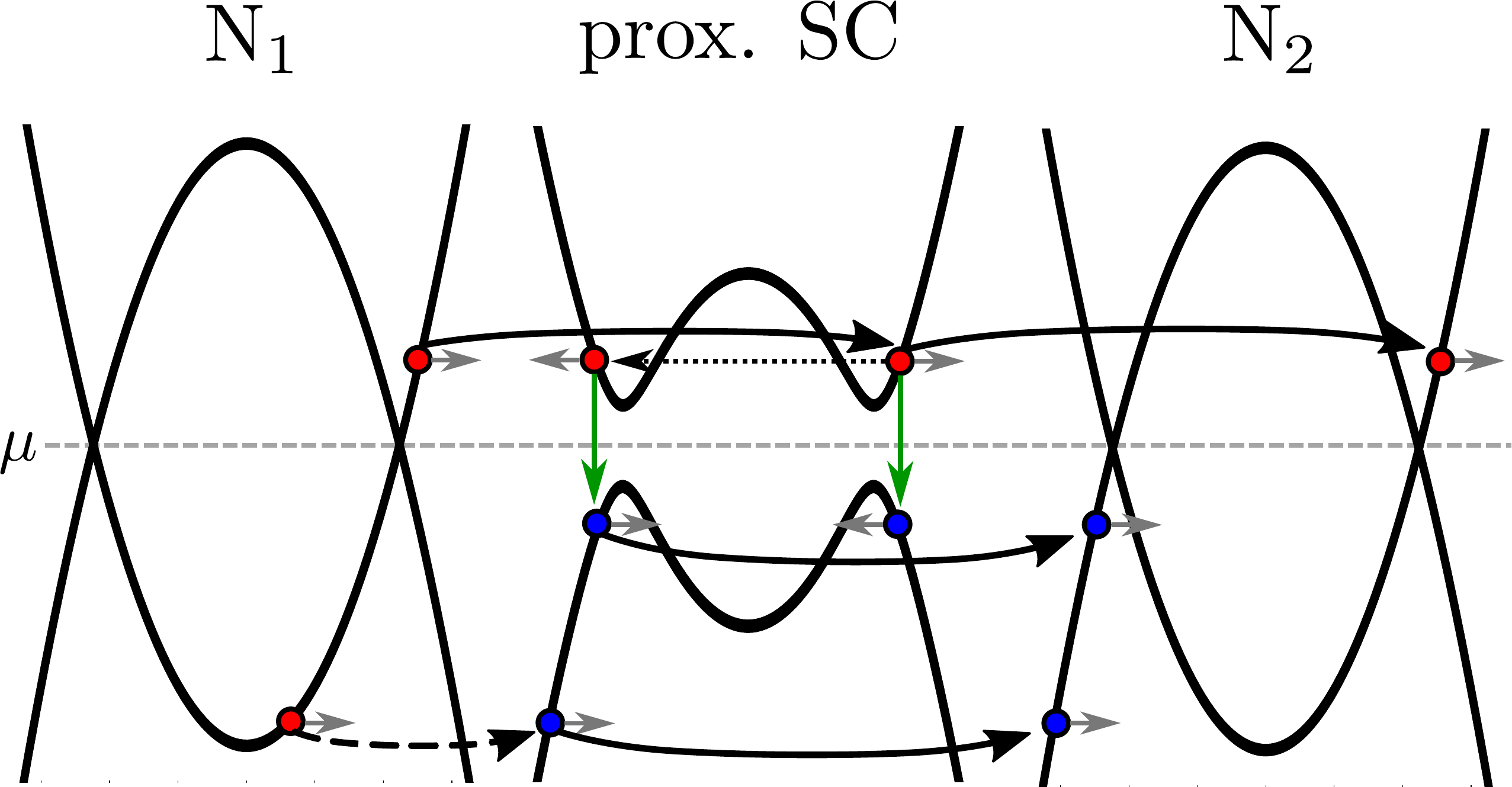}
\caption{Schematic of the quasiparticle transport properties from the normal lead N$_1$, to the lead N$_2$ through the proximitised region. Quasiparticles transferring to a neighboring region (solid black arrows) predominantly preserve the quasiparticle type: electron-like (red dots) or hole-like (blue dots). Andreev reflection (green vertical arrows) changes the quasiparticle type, and the direction of propagation (grey arrows). Disorder scattering (black dotted arrow) changes the propagation direction. Finally, if no quasiparticles of the same type are available, quasiparticle transmission between regions may also result in the change of the quasiparticle type (black dashed arrow).}
\label{fig:dispersions}
\end{figure}

\co{CAR dominated nonlocal conductance is important, however it does not happen under normal circumstances.}
A negative nonlocal conductance, dominated by CAR, is of fundamental interest, since the proximitised system then functions as a Cooper pair splitter \cite{deutscher2000, hofstetter2009, herrmann2010, he2014, chen2015}.
In Sec.~\ref{sec:transport}, we observed that the nonlocal conductance in clean systems at zero magnetic field is generally positive, and a CAR-dominated signal ($G_{21}<0$) is rare. 
The reason for this is shown schematically in Fig.~\ref{fig:dispersions}: an electron entering the proximitised region usually converts into an electron-like quasiparticle. 
Andreev reflection changes both the quasiparticle charge and velocity, so that the resulting hole-like quasiparticle returns to the source.
Therefore under normal circumstances Andreev reflection alone is insufficient to generate a negative nonlocal current.

\begin{figure}[!tbh]
\includegraphics[width=0.97\columnwidth]{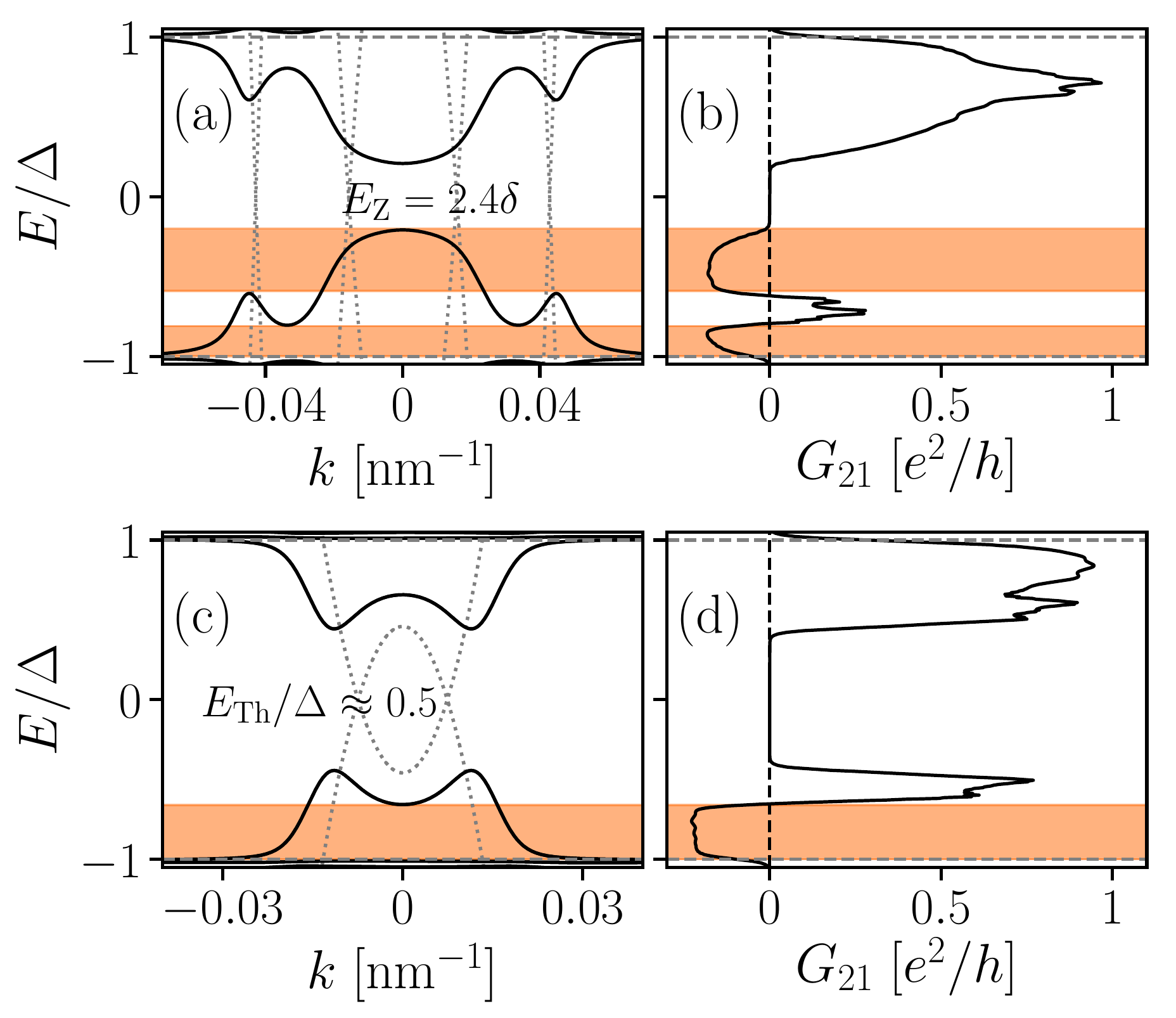}
\caption{ Dispersions (left column) and nonlocal conductance with $L \gg \xi$ (right column) of proximitised channels of width $W = 100$ nm (a, b)  and  $W = 200$ nm (c, d).
Dotted lines show the electron and hole dispersions of the channels with the superconductor removed.
In both cases, the induced gap is smaller than $\Delta$, due to a Zeeman field in (a), and due to $E_{\mathrm{Th}}\lesssim \Delta$ in (c).
There are energy ranges in which only hole-like bands are present, and these correspond to regions of negative $G_{21}$.
Here, (c) is in a low-doping regime $\mu = 0.5$ meV, such that electron modes are absent for $E/\Delta \lesssim -0.5$, producing the hole-like dispersion.
As a result, a larger chemical potential $\mu = 0.8$ meV is needed in the normal leads to observe $G_{21}<0$ at the corresponding energies in (d).
In (a) and (b), we have $\mu = 3$ meV.}
\label{fig:disp_cond_cross}
\end{figure}

\co{In clean systems, CAR dominates when the energy is aligned with hole-like states in the proximitised region.}
Despite $G_{21}$ stays predominantly positive in clean systems, in Sec.~\ref{sec:rectifier} we found that a magnetic field can make the nonlocal conductance negative in large regions of parameter space.
We identify these regions with the presence of only hole-like bands in the proximitized region at the relevant energy, as shown in Fig.~\ref{fig:dispersions}.
If only hole-like states are present in the proximitized region, the incoming electron may only convert into a right-moving hole-like quasiparticle, which in turn converts predominantly into a hole when exiting the proximitized region.
To confirm this argument, we compare the energy ranges where only hole-like quasiparticles are present with the regions of negative $G_{21}$.
Our results are shown in Fig.~\ref{fig:disp_cond_cross}, and they exhibit a very good agreement.
Since the only required property to get a negative nonlocal conductance is a hole-like dispersion relation, this phenomenon does not require SOI, or even Zeeman field. Indeed, our calculations (not shown here) reveal that it is possible to extend the energy ranges over which CAR dominates by filtering the nonlocal conductance by spin, e.g.\ by using magnetically-polarized contacts \cite{falci2001}.

\co{In the low doping-regime, we can engineer a hole-like band structure without a Zeeman field, but to see this manifest as negative nonlocal conductance, a larger chemical potential must be chosen in the normal leads.}
It is possible to systematically obtain a negative nonlocal conductance in the low-doping regime without using a Zeeman field if $\Delta > \Delta_{\mathrm{ind}}$.
This is shown in Fig.~\ref{fig:disp_cond_cross}(c) and (d), where we have also neglected SOI for simplicity.
By choosing $\mu$ comparable to the band offset of the lowest mode in the proximitised channel, at negative energies we obtain an energy range in which the band structure is only hole-like [Fig.~\ref{fig:disp_cond_cross}(c)].
However, the small $\mu$ implies that no electron modes are active in the normal leads in this energy range.
To observe negative nonlocal conductance here, it is therefore necessary to have a larger chemical potential in the normal leads than in the proximitised region, which ensures the presence of propagating electron modes at the relevant energies.
Doing so, we indeed observe a negative nonlocal conductance in the expected energy range of Fig.~\ref{fig:disp_cond_cross}(d).

\co{Disorder does not spoil CAR dominance, and may even be beneficial for CAR, since it enhances the probability of quasiparticle momentum flipping which is needed for CAR.}
Disorder provides an alternative mechanism to obtain negative nonlocal conductance.
Unlike direct electron transfer, which generally conserves the sign of quasiparticle momentum, CAR often requires a sign change of the quasiparticle momentum.
Since disorder breaks momentum conservation, the probabilities of CAR and direct electron transfer become comparable once the system length exceeds the mean free path, and CAR thus more prominent than in a clean system.
Indeed, as shown in Fig.~\ref{fig:nonloc_dis_color}, in disordered systems the nonlocal conductance becomes positive or negative with approximately equal probability.

\section{Summary and outlook}
\label{sec:summary}

\co{Probing induced superconductivity is hard, but we show that nonlocal conductance works for it.}
The standard experimental tool for probing induced superconductivity in a Majorana device is a tunnelling conductance measurement using an attached normal lead.
While this approach detects the density of states, its usefulness is limited because it cannot distinguish the properties in close vicinity of the lead from the properties of the bulk system.
We studied how the \textit{nonlocal} conductance between two spatially separated normal leads attached to the proximitised region overcomes this limitation.
We find that the nonlocal conductance is selectively sensitive to the bulk properties of a proximity superconductor, and allows to directly measure the induced and the bulk superconducting gaps as well as the induced coherence length of the proximitised region.
While we focused on the quasi 1D-systems suitable for the creation of Majorana states, our conclusions are applicable to general proximity superconductors, including 2D materials like graphene covered by a bulk superconductor.

\co{Beyond being able to probe basic properties of proximity superconductivity, nonlocal conductance is sensitive to topological phase transitions, and can result in CAR-dominated response.}
When the probability of CAR is larger than that of electron transmission, the nonlocal conductance turns negative.
While this does not happen normally, we identified conditions that allow CAR to dominate.
This may happen due to disorder, which breaks the relation between quasiparticle charge, velocity and momentum and makes the nonlocal conductance zero on average.
We identified another, systematic way of obtaining dominant CAR by ensuring that the only available states in the proximitised region are hole-like.
A special case of this behaviour is the vicinity of the topological phase transition, where the nonlocal conductance becomes proportional to voltage, resulting in a linear relation between the differential conductance and voltage, or in other words a positive nonlocal current regardless of the sign of the voltage.
This behavior is specific to topological phase transitions, and we showed how it can be used to distinguish accidental low energy states from Majorana states, resolving a potential shortcoming of Majorana tunneling experiments identified in Refs.~\onlinecite{Kells2012,mi2014,Prada2012,Moore2016,liu2017}.

\begin{figure}[!tbh]
\includegraphics[width=0.97\columnwidth]{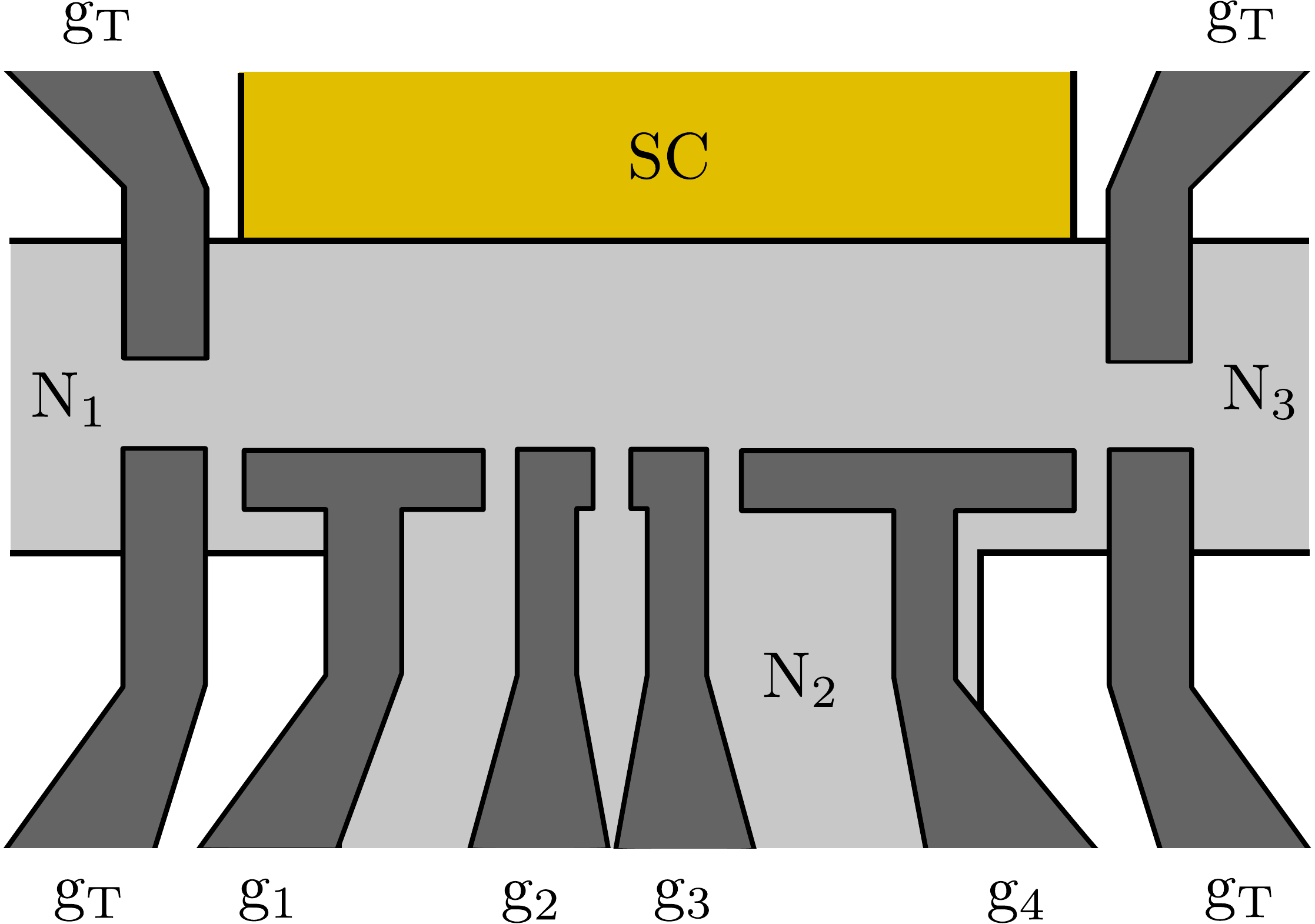}
\caption{A possible experimental realization of a multiterminal proximitized device suited for nonlocal conductance measurements. Electrostatic gates g$_i$, $i \in \{ 1, 2, 3, 4 \}$, pattern out a quasi-one dimensional region in a two-dimensional electron gas, which is proximitised from the side by a superconductor. Gates g$_\text{T}$ create tunnel barriers at the endpoints of the proximised region. Changing the potentials applied to the gates allows for changing the effective device length.}
\label{fig:experimental_setup_gates}
\end{figure}

\co{As a follow-up one could investigate other geometries, Josephson junctions, and better ways to generate CAR by dispersion and geometry.}
Our setup can be used with trivial adjustments to probe the properties of Josephson junctions, proposed as a promising alternative platform for the creation of Majorana states\cite{hell2016, pientka2016}.
Further work could investigate interaction effects on the the nonlocal response \cite{aasen2016}.
An alternative promising avenue of follow-up work is to consider a multiterminal generalization of a nonlocal setup in order to combine local and global sensitivity within the same device.
In Fig.~\ref{fig:experimental_setup_gates} we show a possible experimental realization of such a multiterminal device, where the effective length can be adjusted with gates.
Finally, our results regarding control of the CAR dominance can be used to design devices with a large electron-hole conversion efficiency.

\acknowledgments
We thank D.~Sticlet, M.~P.~Nowak and M.~Wimmer for fruitful discussions. This work was supported by ERC Starting Grant 638760, the Netherlands Organisation for Scientific Research (NWO/OCW), as part of the Frontiers of Nanoscience program and the US Office of Naval Research. MK gratefully acknowledges support from the Carlsberg Foundation.

\appendix
\section{Short, intermediate and long junction limits for hybrid structures}
\label{app:junctions}
\co{We consider the different limits of a one-dimensional junction that consists of a normal part in contact with a much larger superconductor.}
In this appendix, we briefly discuss the subgap spectral characteristics of normal-superconductor junctions in different limits, using heuristic arguments to highlight the essential physics.
For a more rigorous study, we refer the interested reader to e.g.~Refs.~\onlinecite{beenakker1992, volkov1995, pilgram2000, tkachov2005, reeg2016}.
Consider a quasi-one dimensional channel of length $L \rightarrow \infty$ that consists of a junction between a normal part of width $W$ and a superconductor of width $W_{\mathrm{sc}} \gg W$.
The Hamiltonian is the same as in Eq.~\eqref{eq:ham}, but with $p_x \rightarrow \hbar k$ and as before $\Delta \neq 0$ only in the superconductor.
Furthermore, we consider only $E_\text{Z} = 0$ and neglect SOI ($\alpha = 0$) and disorder for simplicity.

\co{The relative size of the superconducting gap and the Thouless energy of the normal part, or equivalently the relative size of the normal part and the BCS coherence length, determines which limit the hybrid system is in.}
The hybrid structure generally has an energy gap $\Delta_{\mathrm{ind}}$, the size of which is determined by two competing energy scales, namely the bulk gap $\Delta$ and the Thouless energy $E_\text{Th} \approx \hbar/ \tau$, with $\tau$ the quasiparticle dwell time in the normal part of the junction.
A short junction has $\Delta \ll E_\text{Th}$ and a long junction $\Delta \gg E_\text{Th}$, while $\Delta \gtrsim E_\text{Th}$ for an intermediate junction.
Alternatively, these conditions are expressed in terms of $W$ and the BCS coherence length $\xi_0 = \hbar v_\text{F} / \Delta$, where $v_\text{F}$ is the Fermi velocity.
For a quasiparticle incident perpendicularly from the normal part to the interface with the superconductor and assuming perfect interface transparency, we have $\tau \propto W/v_\text{F}$ and thus $E_{\mathrm{Th}} \propto \hbar v_\text{F}/W$.
The conditions for short, intermediate and long junctions then become $W \ll \xi_0$, $W \gtrsim \xi_0$ and $W \gg \xi_0$, respectively.
In the short junction limit, we have $\Delta_{\mathrm{ind}} \approx \Delta$, while for long and intermediate junctions $\Delta_{\mathrm{ind}} \propto E_\text{Th}$.

\begin{figure}[!tbh]
\includegraphics[width=0.97\columnwidth]{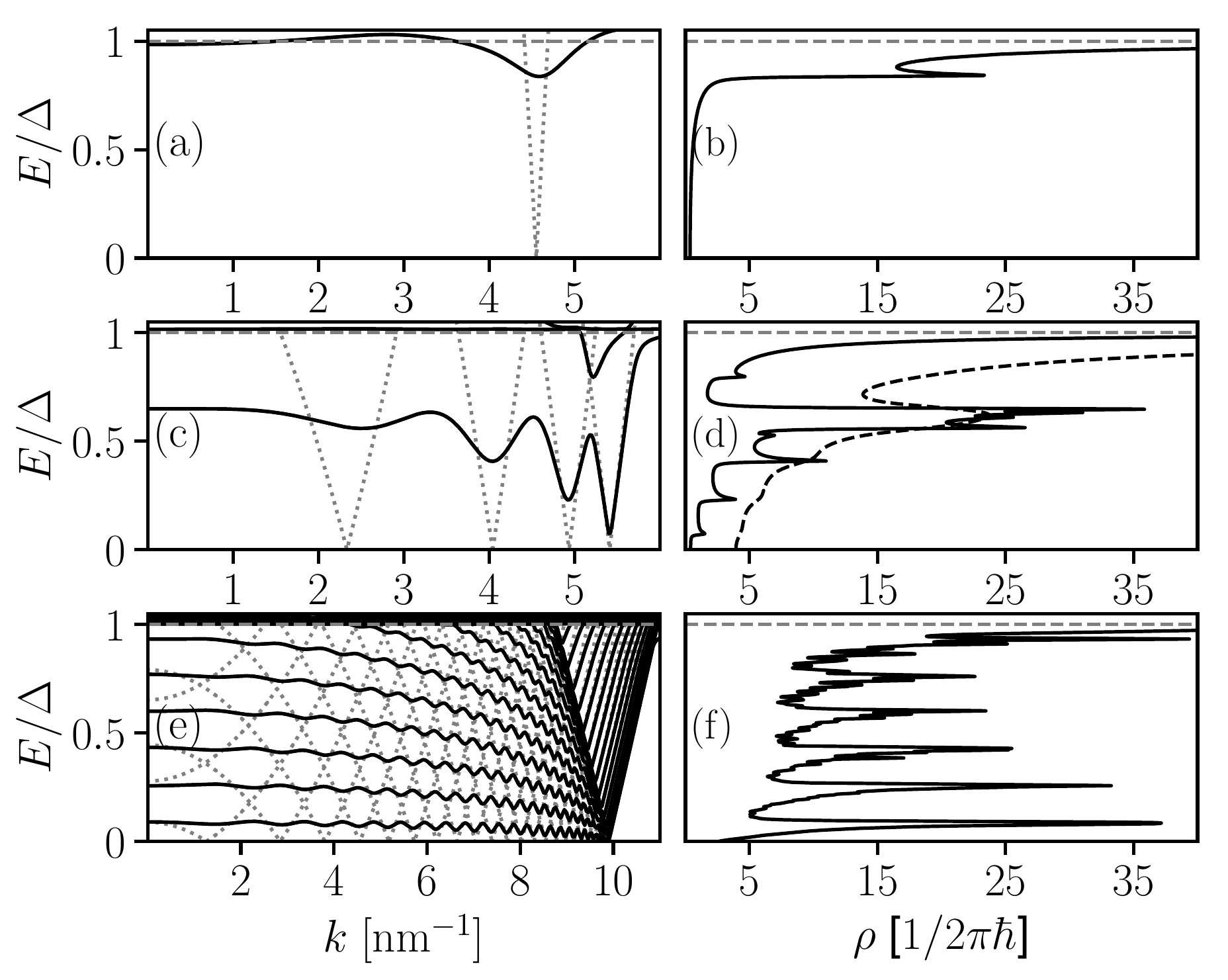}
\caption{Dispersion (left column) and density of states (right column) of a quasi one-dimensional normal-superconductor junction in different regimes: a short junction (top), an intermediate junction (middle) and a long junction (bottom).
In the left column, the dotted curves show the electron and hole dispersions of the corresponding normal channel with the superconductor removed.
In all cases, a small broadening $\Gamma \ll\Delta$ has been added to the density of states.
For the intermediate junction (d), the density of states with larger broadening is also shown (dashed curve).
The curves are symmetric under $(k, E) \rightarrow (\pm k, \pm E)$.}
\label{fig:disp_dos}
\end{figure}

\co{Level spacing in the normal part and the transparency of the interface with the superconductor determine the Thouless energy.}
We now derive a lower bound for $E_\text{Th}$ in terms of the level spacing $\delta$ in the normal part of the junction.
A quasiparticle exiting the superconductor has the dwell time $\tau \propto 2W/\gamma v_{\perp}(k)$ in the normal part.
Here, $v_\perp (k) = \hbar k_\perp(k)/m^*$ and $k_\perp = \sqrt{k_\mathrm{F} - k^2}$ are respectively the velocity and momentum projections perpendicular to the interface with the superconductor at the parallel momentum $k$, with $k_\mathrm{F}$ the Fermi momentum, and $2W$ is the distance the quasiparticle travels before colliding with the superconductor again.
The dwell time scales inversely with the transparency $\gamma$ of the interface between the normal part and the superconductor.
In practice, the transparency is determined by interface properties, such as the presence of a barrier or velocity mismatch, which we parametrize with $0 \leq \gamma \leq 1$ for simplicity.
We thus obtain $E_{\mathrm{Th}}(k) \propto \gamma \hbar^2 \sqrt{k_{\mathrm{F}}^2 - k^2}/2m^* W$.
Observe that $E_{\mathrm{Th}}$ decreases with $k$ and tends to vanish as $k \rightarrow k_\mathrm{F}$ since then $v_\perp \rightarrow 0$.
However, $v_\perp$ is bounded from below in a finite geometry by the momentum uncertainty associated with the band offset, which corresponds to the velocity $\mathrm{d}v_\perp \approx \hbar \pi/m^*W$ in a square-well approximation.
Using $v_\perp = \mathrm{d}v_\perp$ gives the lower bound for the Thouless energy $E_\mathrm{Th} \propto \gamma \hbar^2 \pi/2m^*W^2$.
The preceding discussion implies that in the absence of magnetic fields, the gap in the spectrum of such a junction decreases with momentum to a minimum $\propto 1/m^*W^2$ at $k = k_\mathrm{F}$ [see left column Fig.~\ref{fig:disp_dos}].
Since $\Delta_{\mathrm{ind}}$ is the energy of the lowest Andreev bound state in the junction, we define
\begin{equation} E_\text{Th} = \gamma \delta, \hspace{0.2 cm} \delta = \frac{\hbar^2 \pi^2}{2m^* (2W)^2} \end{equation}
as the Thouless energy of the junction.
Observe that we use $2W$ in the denominator, since that is the distance normal to the interface a quasiparticle travels between successive Andreev reflections \cite{sticlet2016}.

\co{The limit in which the hybrid junction is in strongly affects the dispersion and density of states, with subgap states appearing in the intermediate and long junction regimes.}
The spectral characteristics of a proximitised system strongly depend on which regime the system is in.
Figure \ref{fig:disp_dos} shows the dispersion $\epsilon_n(k)$ and density of states $\rho$ per unit length for junctions in the short, intermediate and long regimes.
The density of states is given by
\begin{equation} \begin{split}
\rho(E) &= \frac{1}{2\pi \hbar} \sum\limits_n \int \delta \left[ E - \epsilon_n(k) \right] \frac{\mathrm{d}E}{\left|v(E) \right|} \\
&= \frac{1}{2\pi \hbar} \sum\limits_n \left| \frac{\mathrm{d}\epsilon_n(k)}{\mathrm{d}k} \right|^{-1}.
\end{split}\end{equation}
Here, $n$ is the subband index including spin and we have used $\hbar v = \mathrm{d}E/\mathrm{d}k$ for the velocity $v$.
In the left column, the solid lines give the dispersion of the hybrid structure, while the dotted lines show the electron and hole dispersions of the normal channel only (with $W_{\mathrm{sc}} = 0$ or $\gamma = 0$).
In all cases, $\mu \gg \Delta$, and $\rho$ has been broadened by convolution with a Lorentzian of full width at half maximum $\Gamma \ll \Delta$.
For the short junction, we indeed have $\Delta_{\mathrm{ind}} \approx \Delta$, which manifests as an essentially hard superconducting gap for $\left| E \right| < \Delta_{\mathrm{ind}}$.
We have verified that $\rho$ vanishes identically in this regime with $\Gamma \rightarrow 0$.
In the intermediate and long regimes, subgap states exist at energies smaller than $\Delta$, which manifests as a nonzero subgap $\rho$ (soft gap).
The difference between the two regimes is the number of these states: in an intermediate junction, they are few, but multiple in the long junction limit, as the conditions $\Delta \gtrsim E_\text{Th}$ and $\Delta \gg E_\text{Th}$ indicate.
Observe that in both cases, the subgap bands are flat around $k = 0$ and drop towards a minimum in energy as $k$ increases before rising sharply again \cite{titov2006-2}.
Superimposed on this are intraband oscillations that happen on a smaller energy scale.
In principle, oscillations thus manifest in $\rho$ on two energy scales: the larger energy scale is the interband spacing around $k = 0$ ($\propto 1/W^2$), and the smaller the scale of intraband oscillations.
Overall, the former has a larger contribution to $\rho$ due to the small curvature in the dispersion.
Oscillations on both scales are clearly visible for the intermediate junction.
However, increasing $\Gamma$ further (dashed curve) washes out the fine structure due to intraband oscillations.
As a result, $\rho$ gradually increases towards a maximum, when $E$ aligns with the energy of the subgap state around $k = 0$.
On the other hand, in the long junction there are multiple states at subgap energies, and the most prominent feature in $\rho$ is the peaks associated with the flat parts of those bands.
The fine structure due to intraband oscillations is superimposed, but masked by the broadening.

\bibliographystyle{apsrev4-1}
\bibliography{nonloc}

\end{document}